\documentclass[journal,draftclsnofoot,onecolumn,12pt]{IEEEtran}
\usepackage{amsthm,amssymb,graphicx,multirow,amsmath,color,amsfonts}
\usepackage{setspace}	
\usepackage[update,prepend]{epstopdf}
\usepackage[noadjust]{cite}
\usepackage[latin1]{inputenc}
\usepackage{tikz}
\usepackage{pdfpages}
\usepackage{subfigure,array}
\usepackage{tabulary}
\usepackage{multirow}
\usepackage{float}
\usepackage{comment}
\usepackage{mathtools}
\let\subparagraph\relax
\usepackage[compact]{titlesec}
\titlespacing{\section}{0pt}{*1}{*1}
\titlespacing{\subsection}{0pt}{*1}{*1}

\include{notation}
\allowdisplaybreaks 
\allowdisplaybreaks

\def\nb0{{\mathbf{0}}}
\def\nb1{{\mathbf{1}}}

\def\ncalB{{\mathcal{B}}}

\def\ncalK{{\mathcal{K}}}

\def\nbbE{{\mathbb{E}}}

\def\nbbP{{\mathbb{P}}}

\def\nbbR{{\mathbb{R}}}

\newtheorem{lemma}{Lemma}

\newtheorem{ndef}{Definition}

\newtheorem{theorem}{Theorem}
\newtheorem{prop}{Proposition}
\newtheorem{cor}{Corollary}

\newtheorem{remark}{Remark}

\def\pc{\mathtt{P_c}}

\def\R{\mathbb{R}}

\def\sinr{\mathtt{SINR}}			

\def\sir{\mathtt{SIR}}

\def\matern{Mat\'ern\ }

\def\caseS{{\sc Type}}
\def\Case{{\sc Type}}
\def\Cases{{\sc Type}s}
\graphicspath{{./Fig/}}
\title{Unified  Analysis of HetNets using Poisson Cluster Process under Max-Power Association}

\author{
Chiranjib Saha, Harpreet S. Dhillon, Naoto Miyoshi, and Jeffrey G. Andrews 
\thanks{C. Saha and H. S. Dhillon are with Wireless@VT, Department of ECE, Virginia Tech, Blacksburg, VA, USA. Email: \{csaha, hdhillon\}@vt.edu. 
N. Miyoshi is with the Department of Mathematical and Computing Science, 
Tokyo Institute of Technology, Tokyo, Japan. Email: miyoshi@is.titech.ac.jp. 
J. G. Andrews is with the Wireless Networking and Communications Group, The University of Texas at Austin, TX, USA. Email: jandrews@ece.utexas.edu. 

The support of the US National Science Foundation (Grant CNS-1617896) and Japan Society for the Promotion of Science (JSPS) Grant-in-Aid for Scientific Research (C) 16K00030 is gratefully acknowledged. 
\hfill
Last updated: \today.
} }

\let\emptyset\varnothing
\begin{document}
\maketitle
\begin{abstract}
Owing to its flexibility in modeling real-world spatial configurations of users and base stations (BSs), the Poisson cluster process (PCP) has recently emerged as an appealing way to model and analyze heterogeneous cellular networks (HetNets). Despite its undisputed relevance to HetNets -- corroborated by the models used in industry -- the PCP's use in performance analysis has been limited. This is primarily because of the lack of analytical tools to characterize performance metrics such as the coverage probability of a user connected to the strongest BS.  In this paper, we develop an analytical framework for the evaluation of the coverage probability, or equivalently the complementary cumulative density function (CCDF) of signal-to-interference-and-noise-ratio ($\sinr$), of a typical user in a $K$-tier HetNet under a $\max$ power-based association strategy, where the BS locations of each tier follow either a Poisson point process (PPP) or a PCP. The key enabling step involves conditioning on the parent PPPs of all the PCPs which allows us to express the coverage probability as a product of sum-product and probability generating functionals (PGFLs) of the parent PPPs. In addition to several useful insights, our analysis provides a rigorous way to study the impact of the cluster size on the ${\tt SINR}$ distribution, which was not possible using existing PPP-based models.
\end{abstract}
\vspace{-.5em}
\begin{IEEEkeywords}
Heterogeneous cellular network,  3GPP, Poisson cluster process, Thomas cluster process, \matern cluster process. 
\end{IEEEkeywords}
\section{Introduction}

Network heterogeneity is at the heart of current 4G and upcoming 5G networks. A key consequence of the heterogeneous deployments is the emergence of different types of spatial couplings across the locations of BSs and users. Perhaps the most prominent one is the {\em user-BS coupling}, where  the users tend to form spatial clusters or {\em hotspots} \cite{saha20173gpp,HetHetNets2015,zhong2018effect} and  small cell BSs (SBSs) are deployed within these hotspots to provide additional capacity. Further, depending on the deployment objectives, the point patterns of BSs of a particular tier may exhibit some {\em intra-tier coupling}, such as clustering patterns for small cells~\cite{3gppreportr12,3gppreportr13} and repulsive patterns for macrocells deployed under a minimum inter-site distance constraint.  Further,  {\em inter-tier coupling} may exist between the locations of BSs of different tiers, for instance, macrocells and small cells when the latter are deployed at the macrocell edge to boost cell edge coverage~\cite{access2010further}.


Not surprisingly, the HetNet simulation models used by the standardization bodies, such as the third generation partnership project (3GPP), are cognizant of the existence of this spatial coupling~\cite{access2010further}. Unfortunately, this is not true for the stochastic geometry based analytical HetNet models, e.g., see~\cite{AndrewsTractable,dhillon2012modeling}, which mostly still rely on the assumption that all network elements (BSs and users) are modeled as independent PPPs. That said, it has been recently shown that PCP-based models are well-suited to capture the aforementioned spatial coupling in a similar way as it is incorporated in 3GPP simulation models~\cite{saha20173gpp,SahaAfshDh2016,Mankar2016,AfshDhiClusterHetNet2016}. Since PCPs are defined in terms of PPPs, they are also very amenable to mathematical analysis. A key prior work in this area is~\cite{saha20173gpp}, which completely characterized the downlink coverage probability for a PCP-based HetNet model under max-$\sinr$  based association scheme in which the typical user connects to the BS offering maximum {\em instantaneous} received $\sinr$~\cite{saha20173gpp}. However, the downlink analysis for the more practical association scheme in which the typical user connects to the BS offering the strongest received power requires a very different mathematical treatment and is still a key open problem. In this paper, we plug this knowledge gap by providing a complete characterization of coverage probability under this association model. 


 \subsection{Background and related works}

Since the use of PPPs for modeling HetNets is by now fairly well-known, we advise interested readers to refer to books, surveys and tutorials, such as \cite{andrews2016primer,elsawy2013stochastic,mukherjee2014analytical,elsawy2016modeling,blaszczyszyn2018stochastic} to learn more about this direction of research. Although sparse, there have been some works on modeling spatial coupling between BSs and users in random spatial models for cellular networks. 
In \cite{NonUniformDhillon},  user-BS coupling was introduced in a PPP-based single tier cellular network model by conditionally thinning the user PPP and biasing user locations towards the BS locations. Owing to the natural connection of the formation of hotspots to the clustering patterns of PCPs, PCP was used to model the user distributions in \cite{DownlinkChiranjib2016,SahaAfshDh2016,Mankar2016,TurgutUEclustering}, where  coupling between the users and BS locations was introduced by placing the SBSs at the cluster centers (parent points) of the user PCP. Further, since intra-tier coupling can be either attractive or repulsive, no single point process is the {\em best} choice for capturing it. For modeling repulsions, \matern hard-core process~\cite{andrews_lte_wifi,parida2017stochastic}, Gauss-Poisson process~\cite{Haenggi_gauss_poisson}, Strauss hardcore process~\cite{PairwiseTaylorDhillon2012},  Ginibre point process~\cite{miyoshi2014cellular,nakata2014spatial},  and more general determinantal point processes~\cite{Li_Dhi_DPP} have been used for BS distributions. On the other hand, for modeling attraction, PCP \cite{AfshDhiClusterHetNet2016} and Geyer saturation process \cite{PairwiseTaylorDhillon2012} have been proposed. For modeling inter-tier coupling, Poisson hole process (PHP) has been a preferred choice~\cite{DengHaenggiHeterogeneous2015,yazdanshenasan2016poisson}, where the macro BSs (MBSs) are modeled as PPP and the SBSs are modeled as another PPP outside the exclusion discs (holes) centered at the MBS locations. Among these ``beyond-PPP'' spatial models of HetNets, the PCP has attracted significant interest because of its generality in modeling variety of user and BS configurations and its mathematical tractability~\cite{saha2017poisson,saha20173gpp}. We now provide an overview of the existing work on  PCP-based models for HetNets. 


In~\cite{SuryaprakashMoller2015,DengHaenggiHeterogeneous2015}, the authors assumed that the SBSs in a two-tier HetNet are distributed as a PCP and derived downlink coverage probability assuming that the serving BS is located at a fixed distance from the receiver, which circumvented the need to consider explicit cell association. While this setting provides useful initial insights, the analysis cannot be directly extended to incorporate realistic cell association rules, such as $\max$-power based association~\cite{jo2012heterogeneous}. The primary challenge in handling cell association in a stochastic geometry setting is to jointly characterize the serving BS distance and the interference field. This challenge does not appear when BS is assumed to lie at a fixed distance from the receiver, such as in \cite{SuryaprakashMoller2015,DengHaenggiHeterogeneous2015}. It is worth noting that the distance of the serving BS from the receiver under $\max$-power based association can be evaluated using contact distance distributions of PCPs, which have recently been characterized in~\cite{AfshSahDhi2016Contact,AfshSahDhi2017ContactMatern}. However, these alone do not suffice because we need to jointly characterize the serving BS distance and the interference field, which is much more challenging. This is the main reason why this problem has remained open for several years.  
In \cite{bao}, the authors characterized handoff rates for a typical user following an arbitrary trajectory in a HetNet with PPP and PCP-distributed BSs. Although the setup is very similar to this paper, the metric considered in \cite{bao} did not require the characterization of coverage probability. 
   In \cite{AzimiAbarghouyi2017StochasticGM}, the authors developed analytical tools to handle this correlation and derived coverage probability when the BSs are modeled as a \matern cluster process (MCP). However, the analysis is dependent on  the geometrical constructions which is very specific to an MCP and is hence not directly applicable to a general PCP.  The coverage analysis for a general PCP was provided in \cite{AfshDhiClusterHetNet2016} where analytical tractability was preserved by assuming that the SBSs to be operating in a {\em closed access} mode, i.e, a user can only connect to a SBS of the same cluster. 
In  \cite{saha20173gpp}, we have provided a comprehensive coverage analysis for the unified HetNet model under $\max$  $\sinr$-based association strategy with $\sinr$ threshold greater than unity. However, the analysis cannot be directly extended to the $\max$ power-based association setup because of the fundamental difference in the two settings from the analytical perspective. 
 Recently, in \cite{miyoshi2018downlink}, the open problem of the coverage analysis for a single tier cellular network under $\max$-power connectivity with PCP-distributed BSs and independent user locations was solved by expressing coverage probability as a sum-product functional over the parent PPP of the BS PCP. Motivated by  this novel analytical approach, in this paper we  provide the complete coverage analysis  for $\max$-power based association strategy  for the unified HetNet model introduced in~\cite{saha20173gpp}.
 \subsection{Contributions}
 We derive the coverage probability of a typical user of a unified $K$-tier HetNet  in which the spatial distributions of $K_1$ BS tiers are modeled as PCPs and $K_2$ BS tiers are modeled as PPPs ($K_1+K_2 = K$). 
 The PCP assumption for the BS tier introduces spatial coupling among the BS locations. We consider two types of users in this network, \Case~1:  users having no spatial coupling with the BSs, and \Case~2: users whose locations are coupled with the BS  locations. For \Case~1, the user locations   
   are modeled as a stationary point process independent of the BS point processes. For \Case~2, the coupling between user and BS
locations  is incorporated by modeling the user locations as a PCP with each user cluster sharing the same cluster center with  a BS cluster. {The  key contributions are highlighted next.}

{\em Exact coverage probability analysis.}
  Assuming that a user connects to the BS offering the maximum average received power, we provide an exact analysis of coverage probability for a typical user which is an arbitrarily selected point from the user point process. 
The key enabler of the coverage probability analysis is  a fundamental property of PCP that conditioned on the parent PPP, the PCP can be viewed as an inhomogeneous PPP, which is a relatively more tractable point process compared to  PCP.
Using this property, we condition on the parent PPPs of all the BS PCPs and derive the conditional coverage probability. Finally, while deconditioning over the parent PPPs, we observe that the coverage probability can be expressed as the product of PGFLs and sum-product functionals of the parent PPPs. 
 This analytical formulation of coverage probability in terms of known point process functionals over the parent PPPs is the key contribution of the paper and yields an easy-to-compute expression of coverage probability under $\max$-power based association. We then specialize the coverage probability for two instances when the PCPs associated with the BSs are either (i) Thomas cluster process (TCP), where the offspring points are normally distributed around the cluster center, or (ii) MCP, where the offspring points are distributed uniformly at random within a disc centered at the cluster center. 
 
{\em System-level insights.}
Using the analytical results, we study the impact of spatial parameters such as cluster size, 
average number of points per cluster and BS density on the coverage probability. Our numerical results demonstrate that  the variation of  coverage probability with cluster size has conflicting trends for \Cases~1 and 2: for \Case~1, coverage decreases as cluster size increases, and for \Case~2, coverage increases as cluster size decreases. As cluster size increases, the coverage probabilities under  \Case~1 and \Case~2 approach the same limit which is the well-known  coverage probability of the PPP-based $K$-tier HetNet~\cite{jo2012heterogeneous}, but from two opposite directions. 
  Our numerical results demonstrate that the impact of the variation of cluster size on coverage probability is not as prominent in \Case~2 as in \Case~1.  

 \section{System Model}
\subsection{PCP Preliminaries}
Before we introduce the proposed PCP-based system model for $K$-tier HetNet, we provide a formal introduction to PCP.
\begin{ndef}[Poisson Cluster Process] \label{def::PCP}
A PCP $\Phi(\lambda_{\rm p}, {g}, \bar{m})$ in $\R^2$ can be  defined as:
\begin{equation}\label{eq::pcp_def}
{\Phi(\lambda_{\rm p},g,\bar{m})= \bigcup_{{\bf z}\in \Phi_{{\rm p}}(\lambda_{\rm p})} {\bf z} + {\cal B}^{\bf z},}
\end{equation} 
where $\Phi_{{\rm p}}=\Phi(\lambda_{\rm p})$ is the parent PPP with intensity $\lambda_{{\rm p}}$ and $\ncalB^{\bf z}$ denotes the offspring point process corresponding to a cluster center at ${\bf z}\in\Phi_{\rm p}$ 
where  $\{{\bf s}\in\ncalB^{\bf z}\}$ is an independently and identically distributed (i.i.d.) sequence of random vectors   with  probability density function (PDF)  {$g({\bf s})$}. The number of points in $\ncalB^{\bf z}$ is denoted by $M$, where $M\sim {\tt Poisson}(\bar{m})$.
\end{ndef}
\begin{table}
\centering
\caption{List of Notations.}\label{tab::notation}
\begin{tabular}{ |c |l| }
\hline
 ${\cal K}$ & Index set of all BS tiers ($|{\cal K}| = K$)\\\hline 
 ${\cal K}_1,{\cal K}_2$ & Index set of all BS tiers modeled as PCP and PPP\\\hline  
$\Phi_k=\Phi(\lambda_{{\rm p}_k},g_k,\bar{m}_k)$ & The point process of the $k^{th}$ BS tier, $k\in{\cal K}_1$, which is  a PCP. \\\hline
 $\Phi_k=\Phi(\lambda_k)$& The point process of the $k^{th}$ BS tier, $k\in{\cal K}_2$, which is a PPP.\\\hline
$P_k$&Transmit power of a BS in $\Phi_k$\\\hline
$\alpha$&Path-loss exponent ($\alpha>2$)\\\hline
$\tau_k$&Coverage threshold of $\Phi_k$\\\hline
$\bar{P}_{j,k}$&$(P_j/P_k)^{1/\alpha}$\\\hline
$f_{{\rm d}_k}(\cdot|z)$&Conditional PDF of distance of a point of $\Phi_k$ ($k\in{\cal K}_1$)\\&from origin given its cluster center is located at ${\bf z}$ ($z=\|{\bf z}\|$)\\\hline
$f_{{c}_k}(\cdot),F_{{c}_k}(\cdot)$ & PDF and CDF of contact distance of $\Phi_k$\\\hline
${\cal C}_{j,k}(r,z)$&$\exp\left(-\bar{m}_k\left(1-\int_{\bar{P}_{j,k}}^{\infty}(1+\frac{\tau_kP_kr^\alpha}{P_k}y^\alpha)^{-1}f_{{\rm d}_j}(y|z){\rm d}y\right)\right)$\\\hline
$\rho(\tau_i,\alpha)$&$ {1+\tau_{i}^{2/\alpha} \int\limits_{\tau_i^{-2/\alpha}}^{\infty} 
\frac{1}{1+t^{\alpha/2}}{\rm d}t =1+ \frac{2\tau_i}{\alpha-2}{}_{2}{\cal F}_{1}\left[1,1-\frac{2}{\alpha};2-\frac{2}{\alpha};-\tau_i\right]}$\\\hline
\end{tabular}
\end{table}
{\bf Notation.} { While we reserve the  symbol $\Phi$ to denote any point process, to indicate whether it is a PCP or PPP  we specify the  parameters in parentheses  accordingly, i.e.,  $\Phi(\lambda_{\rm p },g,\bar{m})$  denotes a  PCP according to Definition~\ref{def::PCP} and $\Phi(\lambda)$  denotes a  PPP with intensity $\lambda$.}
 
A PCP can be viewed as a collection of offspring process ${\ncalB}^{\bf z}$ translated by ${\bf z}$ for each ${\bf z}\in\Phi_{\rm p}$. Then the sequence of points $\{{\bf t}\}\equiv{\bf z}+\ncalB^{\bf z}$ is conditionally i.i.d. with PDF {${f}({\bf t}|{\bf z}) =g({\bf t}-{\bf z})$}. Note that the conditional distribution of the point coordinates given its cluster center at ${\bf z}$ is equivalent to translating a cluster centered at the origin to ${\bf z}$.   
For a PCP, the following result can be established. 
\begin{prop}\label{prop:1}
Conditioned on the parent point process $\Phi_{\rm p}$, $\Phi(\lambda_{\rm p},g,\bar{m})$ is an inhomogeneous PPP with intensity 
\begin{equation}\label{eq::intensity}
\lambda({\bf x})=\bar{m}\sum_{{\bf z}\in\Phi_{\rm p}}f({\bf x}|{\bf z}).
\end{equation}
\end{prop}
\begin{IEEEproof}{
While one can prove this result for a more general setting of Cox processes (see~\cite{chiu2013stochastic}), we prove this result for PCP for completeness as follows.} 
Let $N_{\Phi}$ be the random counting measure associated with the  point process $\Phi$. Then, {for  a Borel set $A\in B(\R^2)$, where $B(\R^2)$ is the Borel $\sigma$-algebra on $\R^2$}, $N_\Phi(A)$ is a random variable denoting the number of points of $\Phi$ falling in $A$.  First it is observed that for ${\cal B}^{\bf z}$, $N_{{\cal B}^{\bf z}}(A)\sim{\tt Poisson}(\bar{m}\int_{A}f({\bf y}|{\bf z}){\rm d}{\bf y})$ since the probability generating function (PGF) of $N_{{\cal B}^{\bf z}}(A)$ is 
\begin{align*}
&\nbbE\left[\theta^{N_{{\cal B}^{\bf z}}(A)}\right] = \nbbE\left[\theta^{\sum_{i=1}^{M}{\bf 1}({\bf s}_i\in A)}\right] = \nbbE\left[\prod_{i=1}^M\theta^{{\bf 1}({\bf s}_i\in A)}\right] \stackrel{(a)}{=} \nbbE\left[\prod_{i=1}^M\nbbE\left[\theta^{{\bf 1}({\bf s}_i\in A)}\right]\right] \\&\stackrel{(b)}{=}\nbbE\left[\prod_{i=1}^M 
\left(1-(1-\theta)\int_Af({\bf y}|{\bf z}){\rm d}{\bf y}\right)\right] =  \nbbE\left[\left(1-(1-\theta)\int_Af({\bf y}|{\bf z}){\rm d}{\bf y})\right)^M\right]\\
&\stackrel{(c)}{=} \exp\left(-\bar{m}\int_Af({\bf y}|{\bf z}){\rm d}{\bf y}\left(1-\theta\right)\right).
\end{align*}
 Here $(a)$ follows from the fact that  the offspring points are i.i.d. around the cluster center at $\bf z$, $(b)$ and $(c)$ are obtained by using the PGF of Bernoulli and Poisson distributions, respectively. Hence $N_{{\cal B}^{\bf z}}(A)\sim{\tt Poisson}(\bar{m}\int_Af({\bf y}|{\bf z}){\rm d}{\bf y})$.   
Now, conditioned on $\Phi_{\rm p}$ the PGF of $N_{\Phi}(A)$ is expressed as: 
\begin{align*}
&\nbbE[\theta^{N_\Phi(A)}|\Phi_{\rm p}] = \nbbE\left[\theta^{\sum\limits_{{\bf z}\in\Phi_{\rm p}}N_{{\cal B}^{\bf z}}(A)}\bigg|\Phi_{\rm p}\right]= \nbbE\left[\prod\limits_{{\bf z}\in\Phi_{\rm p}}\theta^{N_{{\cal B}^{\bf z}}(A)}|\Phi_{\rm p}\right] \stackrel{(a)}{=} \prod\limits_{{\bf z}\in\Phi_{\rm p}}\nbbE\left[\theta^{N_{{\cal B}^{\bf z}}(A)}|\Phi_{\rm p}\right]\\& \stackrel{(b)}{=}\prod\limits_{{\bf z}\in\Phi_{\rm p}}\exp\left(-\bar{m}\int_{A}f({\bf y}|{\bf z}){\rm d}{\bf y}(1-\theta)\right)=\exp\left(-\bar{m}\sum\limits_{{\bf z}\in\Phi_{\rm p}}\int_{A}f({\bf y}|{\bf z}){\rm d}{\bf y}(1-\theta)\right),
\end{align*}
where $(a)$ follows from the fact that conditioned on $\Phi_{\rm p}$, $\{{\cal B}^{\bf z}\}$ is sequence of i.i.d. offspring point processes, $(b)$ is obtained by substituting the PGF of $N_{{\cal B}^{\bf z}}(A)$. Hence it is observed that $N_\Phi(A)|\Phi_{\rm p}\sim{\tt Poisson}\left(\bar{m}\sum_{{\bf z}\in\Phi_{\rm p}}\int_A f({\bf y}|{\bf z}){\rm d}{\bf y}\right)$. Thus $\Phi|\Phi_{\rm p}$ is an inhomogeneous PPP with intensity measure $\Lambda(A) = \bar{m}\sum_{{\bf z}\in\Phi_{\rm p}}\int_A f({\bf y}|{\bf z}){\rm d}{\bf y}$. 
\end{IEEEproof}
  \begin{figure}
          \centering
          \subfigure[\Case~1]{
              \includegraphics[width=.35\linewidth]{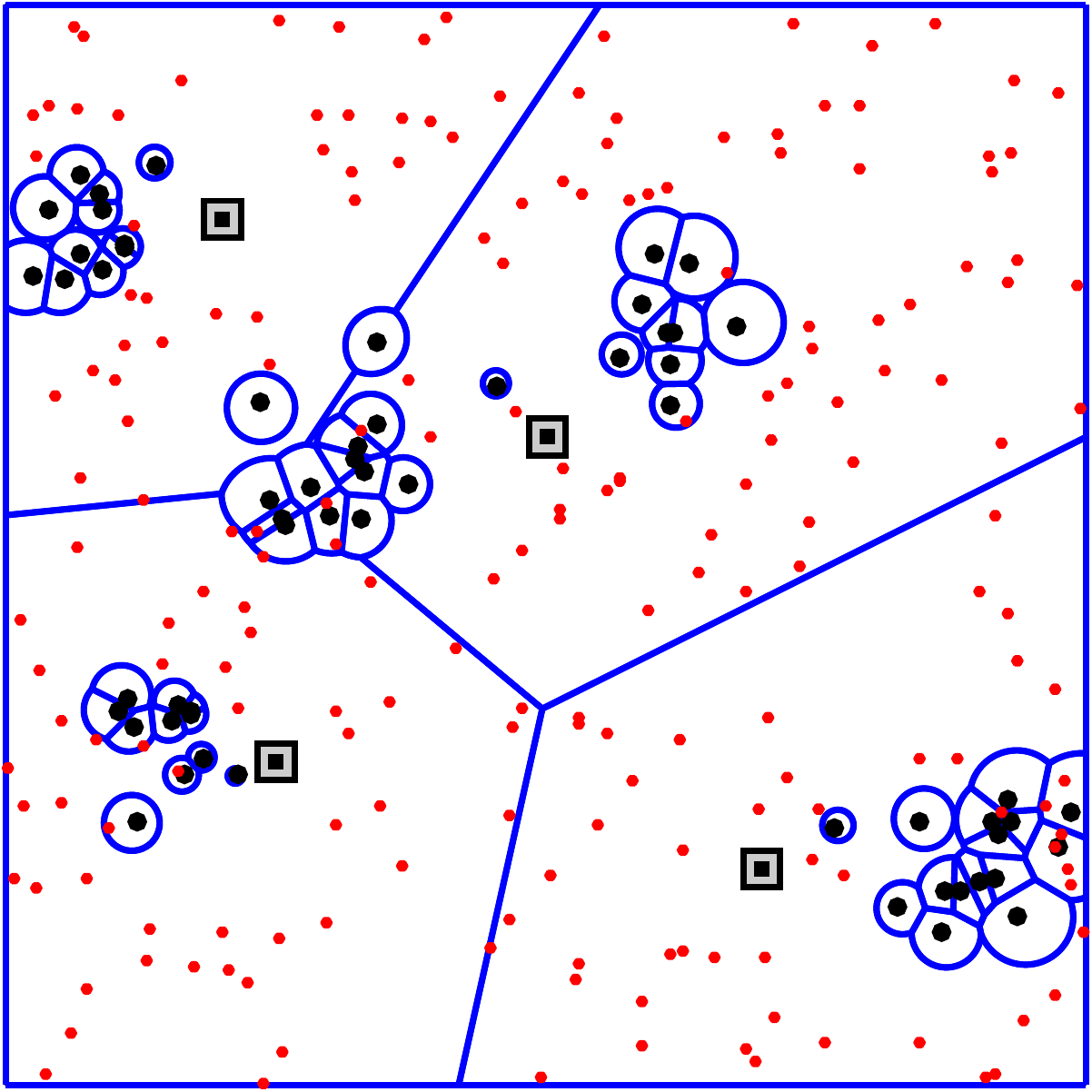}
               \label{fig::system::model::case1}
               }
          \subfigure[\Case~2]{
              \includegraphics[width=.35\linewidth]{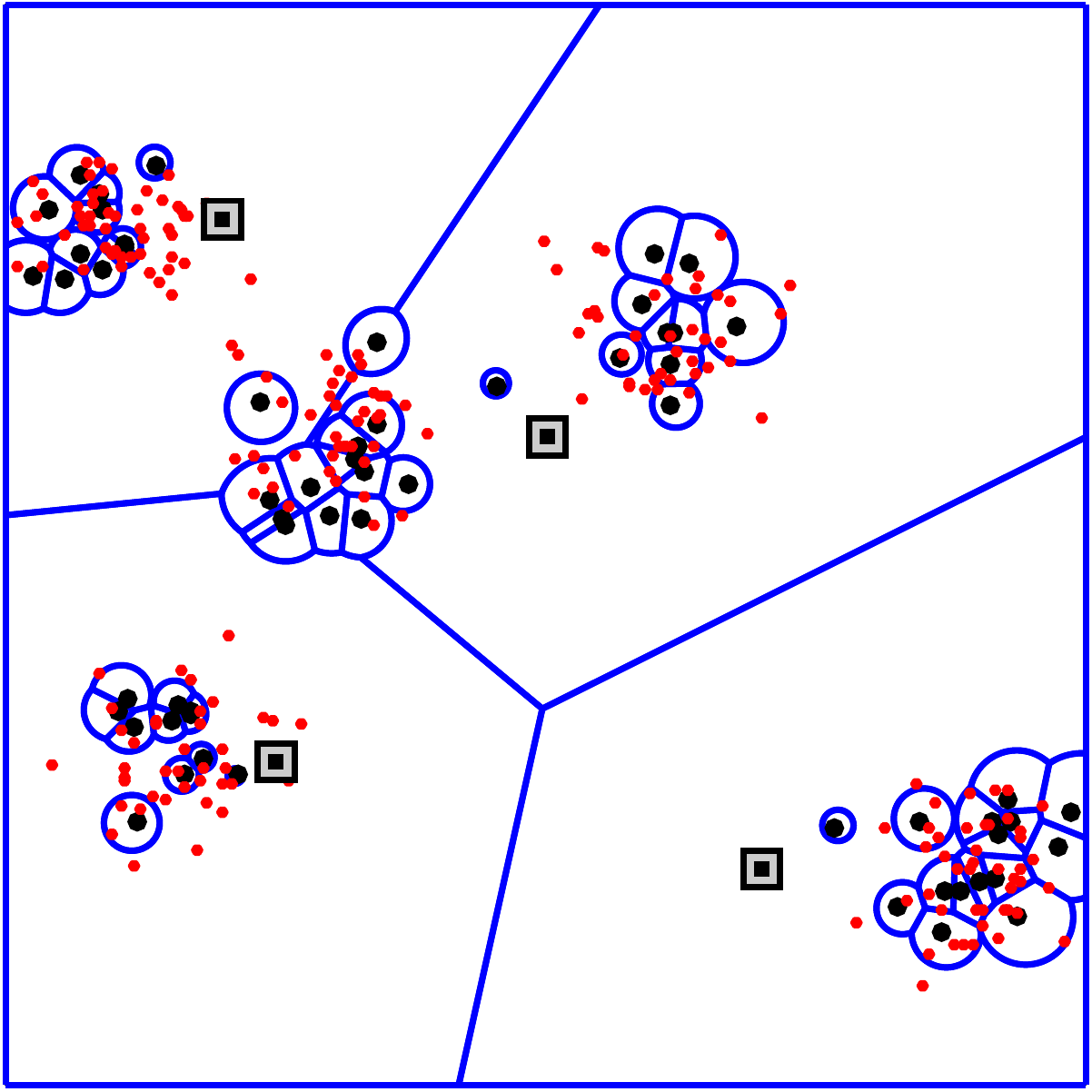}
              \label{fig::system::model:case2}             
              }
              \caption{An illustration of the a two-tier HetNet, where $\Phi_1$ is a PCP of SBSs (illustrated as black dots) and $\Phi_2$ is a PPP of MBSs (illustrated as squares). The users (points of $\Phi_{\rm u}$) are illustrated as red dots. In  \subref{fig::system::model::case1}, $\Phi_{\rm u}$ is a PPP, and in \subref{fig::system::model:case2}, $\Phi_{\rm u}$ is a PCP with the same parent PPP as that of $\Phi_1$.}
             \label{fig::system::model}
          \end{figure} 
\subsection{$K$-tier HetNet Model}\label{subsec::k-tier-hetnet}
We assume a $K$-tier HetNet where BSs of each tier are distributed as a  PPP or PCP. Let $\ncalK_1$ and  $\ncalK_2$ denote the index sets  of the BS  tiers which are modeled as PCP and PPP,  respectively, with  $|\ncalK_1\cup\ncalK_2| = K$ and $\ncalK_1\cap\ncalK_2 = \emptyset$. 
We denote the point process of the $k^{th}$ BS tier as $\Phi_k$,  where $\Phi_k$ is either a  PCP i.e. $\Phi(\lambda_{{\rm p}_k},g_k,\bar{m}_k)$ ($\forall k\in\ncalK_1$) where $\Phi_{{\rm p}_k}=\Phi(\lambda_{{\rm p}_k})$ is the parent PPP  
or a PPP $\Phi(\lambda_k)$ ($\forall  k\in\ncalK_2$). Also define $f_k({\bf x}|{\bf z})={g}_k({\bf x}-{\bf z}),\ \forall\ k\in{\cal K}_1$.  
Each BS of $\Phi_k$ transmits at constant power $P_k$. 
 We assume that the users are distributed according to a stationary point process $\Phi_{\rm u}$. 
 We now consider two types of users in the network. 
 
 \caseS~1: {\em No user-BS coupling.} The first type of users are uniformly distributed over the network, such as, the pedestrians and users in transit, and their locations are independent of the BS locations.  {There is no restriction on the distribution of these users as long as the distribution is stationary. For instance, one way of modeling the locations of these users is to assume that they are distributed as a  homogeneous  PPP.} 
 
 \caseS~2: {\em User-BS coupling.} The second type of users are assumed to form spatial clusters (also called {\em user hotspots})  and their locations can be modeled as a PCP $\Phi_{\rm u} = \Phi (\lambda_{{\rm p}_u},g_{\rm u},\bar{m}_{\rm u})$~\cite{SahaAfshDh2016,saha20173gpp}. When the users are clustered, we also assume that one BS tier (say, the $q^{th}$ tier, $q\in{\cal K}_1$) is deployed to serve the user hotspots, thus introducing  coupling between  ${\Phi}_{\rm u}$ and $\Phi_{q}$.  In other words, $\Phi_{\rm u}$ and $\Phi_{q}$ are two PCPs having same parent PPP $\Phi_{{\rm p}_ q}\equiv\Phi_{{\rm p}_{\rm u}}$. 
 {Hence conditioned on $\Phi_{{\rm p}_q}$, $\Phi_{\rm u}$ and $\Phi_q$ are (conditionally) independent but not identically distributed.}  
 This assumption is motivated by the way SBSs are placed at higher densities in the locations of user hotspots in 3GPP simulation models of HetNets~\cite{3gppreportr12,3gppreportr13}. 
 
\begin{remark}For \caseS~1, $\Phi_{\rm u}$ can be any general stationary point process including PPP and for \caseS~2, we do not specify   ${\bar{m}}_{\rm u}$  of $\Phi_{\rm u}$, since $\bar{m}_{\rm u}$ does not appear explicitly in the coverage analysis. However, these specifications of $\Phi_{\rm u}$ are required when one has to characterize other metrics like BS load and rate coverage probability~\cite{LoadAwareDhillon2013,OffloadingSingh}. Further,   the coverage probability analysis that follows can be extended for  a general user distribution  which is the superposition of PPP and PCPs along similar lines to \cite{SahaAfshDh2016}.
\end{remark}
 In Fig.~\ref{fig::system::model}, we provide an illustration of the system model. 
 For both  types of user distributions, we perform our analysis for a {\em typical} user which corresponds to a  point selected uniformly at random from $\Phi_{\rm u}$. Since $\Phi_{\rm u}$ is stationary, the typical user is assumed to be located at the origin without loss of generality.  
%
 For \caseS~1, since $\Phi_{\rm u}$ and $\Phi_{k}$, $\forall\ k\in{\cal K}$ are independent, the selection of the typical user does not bias the distribution of $\Phi_k$. 
 However, for  \caseS~2, the selection of the typical user affects the BS point process $\Phi_q$ due to the existence of user-BS coupling.  We assume that  the typical user belongs to a cluster centered at ${\bf z}_0$. By construction, $\Phi_{q}$ is always conditioned to have a cluster centered at ${\bf z}_0$.  Hence, the typical user will {\em see} the palm version of $\Phi_q$ which, by Slivnyak's theorem, is equivalent to 
the superposition of $\Phi_{q}$ and ${\bf z}_0+{\cal B}^{{\bf z}_0}$ where $\Phi_{q}$ and ${\bf z}_0+{\cal B}^{{\bf z}_0}$ are independent. For \caseS~2, we modify $\Phi_q $ as $\Phi_q = \Phi(\lambda_{{\rm p}_q},g_q,\bar{m}_q)\cup{\bf z}_0+{\cal B}^{{\bf z}_0}$. Consequently, 
 the underlying parent point process  is modified as $\Phi_{{\rm p}_q} = \Phi(\lambda_{{\rm p}_{q}})\cup \{{\bf z}_0\}$. The BS cluster ${\cal B}^{{\bf z}_0}$ is termed as the {\em representative cluster}.
 
Considering the typical user at origin, the downlink power received from a BS at ${\bf x}\in\Phi_k$ is expressed as $P_{k}h_{\bf x}\|{\bf x}\|^{-\alpha}$, where $h_{\bf x}$ and $\alpha$ denote the small scale fading and path-loss exponent, respectively. 
 We assume that each link undergoes  Rayleigh fading, hence $\{h_{\bf x}\}$ is a sequence of i.i.d. random variables with $h_{\bf x}\sim \exp(1)$.  
The user connects to the BS providing the maximum received power averaged over fading. Thus, if ${\bf x}^*$ is the location of the serving BS, then, ${\bf x}^* = \arg\max_{\{\widetilde{\bf x}_k,k\in{\cal K}\}}P_k{\|\widetilde{\bf x}_k\|}^{-\alpha}$, where $\widetilde{\bf x}_k = \arg\max_{{\bf x}\in\Phi_k}P_k\|{\bf x}_k\|^{-\alpha} = \arg\min_{{\bf x}\in\Phi_k}\|{\bf x}\|$ is the location of candidate serving BS in $\Phi_k$. {Note that the candidate serving BS in $\Phi_k$ 
is the BS in $\Phi_k$ which is geometrically closest  to the user.} 
We first define the {\em association event} corresponding to the $i^{th}$ tier as the event that the serving BS belongs to $\Phi_i$, denoted as ${\cal S}_i =\{ {\bf x}^* = \widetilde{{\bf x}_i}\}$. Conditioned on ${\cal S}_i$,  the $\sinr$ experienced by the typical user is 
\begin{equation}\label{eq::sinr}
\sinr({\bf x}^*) = \frac{P_i h_{{\bf x}^*}\|{\bf x}^*\|^{-\alpha}}{{\tt N}_0+\sum\limits_{j\in{\cal K}}\sum_{{{\bf x}\in\Phi_j\setminus\{{\bf x}^*\}}}P_jh_{\bf x}\|{\bf x}\|^{-\alpha}},
\end{equation}    
where ${\tt N}_0$ is the thermal noise power. We define the coverage probability as the probability of the union of $K$ mutually exclusive coverage events 
\begin{equation}\label{eq::coverage:definition}
\pc = \nbbP\left(\bigcup\limits_{i\in{\cal K}}\{\sinr({\bf x}^*)>\tau_i,{\cal S}_i\}\right) = \sum\limits_{i\in{\cal K}}\nbbP\left(
\sinr({\bf x}^*)>\tau_i,{\cal S}_i\right),
\end{equation} 
where we call the $i^{th}$ term under the summation as the {\em $i^{th}$ tier  coverage probability} which is the joint probability of the  events ${\cal S}_i$ and $\{\sinr({\bf x}^*)>\tau_i\}$. Here $\tau_i$ is the $\sinr$ threshold for the $i^{th}$ tier required for successful demodulation and decoding of the received signal.

\section{Coverage Probability Analysis}
 {
We begin our   coverage analysis by first conditioning on every parent PPP, i.e., $\Phi_{{\rm p}_k},\ \forall\ k\in{\cal K}_1$.}  Following Proposition~\ref{prop:1}, $\Phi_{k}|\Phi_{{\rm p}_k}$ will be inhomogeneous PPP with intensity $\lambda_k({\bf x})=\bar{m}_k\sum_{{\bf z}\in \Phi_{{\rm p}_k}}f_k({\bf x}|{\bf z})$. A slightly different situation occurs for $\Phi_{q}$ in \caseS~2. However,   once $\Phi_{{\rm p}_q}$ is modified by adding a point at ${\bf z}_0$ to the original parent PPP,  $\Phi_{q}|\Phi_{{\rm p}_q}$ again becomes an inhomogeneous PPP with intensity $\bar{m}_q\sum_{{\bf z}\in\Phi_{{\rm p}_q}\cup\{{\bf z}_0\}}f_q({\bf x}|{\bf z})$.
 
 \textbf{Notation.} To denote the  distance of a point ${\bf y}\in \R^2$, we use $y$ and $\|{\bf y}\|$ interchangeably. 
 \subsection{Contact Distance Distribution}
 First we will derive the  distribution of $\|\widetilde{\bf x}_k\|$, $k\in{\cal K}$, or the distance distribution of the candidate serving BS of tier $k$. Since $\|\widetilde{\bf x}_k\|$ is the nearest BS to the typical user which is at origin,  the distribution of  $\|\widetilde{\bf x}_k\|$ is the  same as the contact distance distribution of $\Phi_k$, denoted as $f_{c_k}(r|\Phi_{{\rm p}_k})$.  Let $f_{{\rm d}_k}(r|{\bf z})$ and $F_{{\rm d}_k}(r|{\bf z})$ denote the PDF and CDF of the distance of a randomly selected point of $\Phi_k$ ($k\in{\cal K}_1$) given its cluster center is located at ${\bf z}$.  
 Before presenting the contact distance distributions, we observe the following property of the conditional distance distribution.
\begin{lemma}\label{lemm::vector::independence}
If the offspring points are isotropically distributed around the cluster center i.e., the radial coordinates of the offspring points with respect to the cluster center  have the joint PDF  $g_{k}(s,\theta_s) = g_{k(1)}(s)\frac{1}{2\pi}$, where $g_{k(1)}(\cdot)$ is the marginal PDF of the radial  coordinate, then,  
$f_{{\rm d}_k}(r|{\bf z}) =f_{{\rm d}_k}(r|{z})  $ and  $F_{{\rm d}_k}(r|{\bf z})=F_{{\rm d}_k}(r|{z})$. That is, the conditional distance distribution depends only on the magnitude of ${\bf z}$. 
\end{lemma} 
\begin{IEEEproof}Let ${\bf x}\in\Phi_{k}$ is the location of the point whose cluster center is located at ${\bf z}$. Then, 
\begin{align*} 
F_{{\rm d}_k}(r|{\bf z})&=\nbbP(\|{\bf x}\|<r|{\bf z})=\int\limits_{b({\bf 0},r)} g_k({\bf x}-{\bf z}){\rm d}{\bf x}\\ 
&= \int\limits_{0}^{2\pi}\frac{1}{2\pi}\int\limits_{0}^{r}g_{k(1)}(\sqrt{x^2+z^2-2xz\cos(\theta_x-\theta_z)})x{\rm d}x{\rm d}\theta_x \\&=\int\limits_{\theta_z}^{2\pi+\theta_z}\frac{1}{2\pi}\int\limits_{0}^{r}g_{k(1)}(\sqrt{x^2+z^2-2xz\cos\theta})x{\rm d}x{\rm d}\theta.
\end{align*}
Here $b({\bf 0},r)$ denotes a disc of radius $r$ centered at the origin. Differentiating with respect to $r$,
\begin{equation}\label{eq::distance::dist::intermediate}
f_{{\rm d}_k}(r|{\bf z}) = \int\limits_{\theta_z}^{2\pi+\theta_z}\frac{1}{2\pi}g_{k(1)}(\sqrt{r^2+z^2-2rz\cos\theta})r{\rm d}\theta.
\end{equation}
Since the region of the  integral over $\theta$ is the perimeter of the disc $b({\bf 0},z)$, it is independent of the choice of $\theta_z$.
\end{IEEEproof}
We now characterize the PDF of $\|\widetilde{{\bf x}}_k\|$.
 \begin{lemma}\label{lemm::contact::distance::PCP}
 For $k\in{\cal K}_1$, conditioned on $\Phi_{{\rm p}_k}$, the PDF and CDF of $\|\widetilde{\bf x}_k\|$ are given as:
 \begin{equation}\label{eq::contact::distance::PCP}
 f_{c_k}(r|\Phi_{{\rm p}_k})= \bar{m}_k\sum\limits_{{\bf z}\in\Phi_{{\rm p}_k}} f_{{\rm d}_k}(r|z)\prod\limits_{{\bf z}\in\Phi_{{\rm p}_k}} \exp\left(-\bar{m}_k F_{{\rm d}_k}(r|z)\right),\qquad r\geq 0,
 \end{equation}
 and
 \begin{equation}
 F_{c_k}(r|\Phi_{{\rm p}_k}) = 1-\exp\bigg(- \bar{m}_k\sum\limits_{{\bf z}\in\Phi_{{\rm p}_k}}\int\limits_{0}^{r}f_{{\rm d}_k}( y|{ z}){\rm d}{ y}\bigg),\qquad r\geq 0.
 \end{equation}
 \end{lemma}
 \begin{IEEEproof}
 For $k\in{\cal K}_1$, the CDF of $\|\widetilde{\bf x}_k\|$ is
\begin{align*}
F_{c_k}(r|\Phi_{{\rm p}_k}) &= 1-\nbbP(\Phi(b({\bf 0},r))=0|\Phi_{{\rm p}_k}) \\&= 1-\exp\bigg(-\int_{b({\bf 0},r)}\lambda_k({\bf x}){\rm d}{\bf x}\bigg)\\& =1-\exp\bigg(- \bar{m}_k\sum\limits_{{\bf z}\in\Phi_{{\rm  p}_k}}\int\limits_{b({\bf 0},r)}f_k({\bf y}|{\bf z}){\rm d}{\bf y}\bigg)\\
&=
1-\exp\bigg(- \bar{m}_k\sum\limits_{{\bf z}\in\Phi_{{\rm p}_k}}\int\limits_{0}^{r}f_{{\rm d}_k}( y|{ z}){\rm d}{ y}\bigg).
\end{align*}
The last step is due  the fact that integrating a joint PDF of polar coordinates over $b({\bf 0},r)$ is equivalent to integrating the marginal PDF of radial coordinate over $(0,r]$. The final result in \eqref{eq::contact::distance::PCP} can be obtained by differentiating the CDF with respect to $r$.
 \end{IEEEproof}
Note that  one can obtain the PDF of the  contact distance of PCP by deconditioning $f_{c_k}(r|\Phi_{{\rm p}_k})$ over $\Phi_{{\rm p}_k}$~\cite{miyoshi2018downlink}. This is an alternative approach to the one  presented in~\cite{AfshSahDhi2016Contact,AfshSahDhi2017ContactMatern} for the derivation of contact distance distribution of PCP.
 
 When $\Phi_k$ is a PPP, i.e., $k\in{\cal K}_2$, the distribution of $\|\widetilde{{\bf x}}_k\|$ is the well-known Rayleigh distribution, given by:
 \begin{equation}\label{eq::contact::distance::PPP}
 f_{c_k}(r) = 2\pi\lambda_{k}r\exp(-\pi\lambda_k r^2),\ F_{c_k}(r) =1-\exp(-\pi\lambda_k r^2), \ r\geq 0.
 \end{equation}
\subsection{Association Probability and Serving Distance Distribution}
We define association probability to the  $i^{th}$ tier as $\nbbP({\cal S}_i)$. The association probability is derived as follows. 
\begin{lemma}\label{lemm::association::probability::PCP}Conditioned on $\Phi_{{\rm p}_k},\forall\ k\in{\cal K}_1$, the association probability to the $i^{th}$ tier is given by: 
$\nbbP({\cal S}_i |\Phi_{{\rm p}_k},\forall\ k\in{\cal K}_1)= $
\begin{subequations}\label{eq::association::probability}
 \begin{multline}
\bar{m}_i\int_{0}^{\infty}\sum\limits_{{\bf z}\in\Phi_{{\rm p}_i}}f_{{\rm d}_i}(r|{z}) \prod_{j_1\in{\cal K}_1}\prod\limits_{{\bf z}\in\Phi_{{\rm p}_{j_1}}} \exp\left(-\bar{m}_{j_1}F_{{\rm d}_{j_1}}\left(\bar{P}_{j_1,i}r|{z}\right)\right)\prod\limits_{j_2\in{\cal K}_2} \exp\left(-\pi\lambda_{j_2}\bar{P}_{j_2,i}^2r^2\right){\rm d}r, \\\text{if }i\in{\cal K}_1,
\end{multline}
\begin{equation}
2\pi\lambda_i\int_{0}^{\infty}\prod\limits_{j_1\in{\cal K}_1}\prod\limits_{{\bf z}\in\Phi_{{\rm p}_{j_1}}} \exp\left(-\bar{m}_{j_1}F_{{\rm d}_{j_1}}\left(\bar{P}_{j_1,i}r|{z}\right)\right) \prod\limits_{j_2\in{\cal K}_2}\exp\left(-\pi\lambda_{j_2}\bar{P}_{j_2,i}^2r^2\right)r{\rm d}r,\text{if }i\in{\cal K}_2 ,
\end{equation}
\end{subequations}
where $\bar{P}_{j,i}= (P_j/P_i)^{1/\alpha}$. 
\end{lemma}
\begin{IEEEproof}
See Appendix~\ref{app::association::probability::PCP}.
\end{IEEEproof}
We  now derive the PDF of the conditional serving distance i.e. $\|{\bf x}^*\|$ given ${\cal S}_i$ and ${\Phi}_{{\rm p}_k}, \forall\ k\in{\cal K}_1$. 
\begin{lemma}\label{lemm::serving::distance::distribution}
The PDF of $\|{\bf x}^*\|$ conditioned on association to the $i^{th}$ tier and $\Phi_{{\rm p}_{k}},\forall\ k\in{\cal K}_1$  is given as: $f_{s_i}(r|{\cal S}_i,\Phi_{{\rm p}_k},\forall\ k\in{\cal K}_1) = $
\begin{subequations}\label{eq::serving::distance::distribution}
\begin{multline}\label{eq::serving::distance::distribution::PCP}
\frac{\bar{m}_i}{\nbbP({\cal S}_i |\Phi_{{\rm p}_k},\forall\ k\in{\cal K}_1)} \sum\limits_{{\bf z}\in\Phi_{{\rm p}_i}}f_{{\rm d}_i}(r|{\bf z}) \prod\limits_{j_1\in{\cal K}_1}\prod\limits_{{\bf z}\in\Phi_{{\rm p}_{j_1}}} \exp\left(-\bar{m}_{j_1}F_{{\rm d}_{j_1}}\left(\bar{P}_{j_1,i}r|{ z}\right)\right)\times\\
\prod\limits_{j_2\in{\cal K}_2}\exp\left(-\pi\lambda_{j_2}\bar{P}_{j_2,i}^2r^2\right),\  r\geq 0,\text{if }i\in{\cal K}_1,
\end{multline}
\begin{multline}\label{eq::serving::distance::distribution::PPP}
\frac{2\pi\lambda_i}{\nbbP({\cal S}_i |\Phi_{{\rm p}_k},\forall\ k\in{\cal K}_1)}\prod\limits_{j_1\in{\cal K}_1}\prod\limits_{{\bf z}\in\Phi_{{\rm p}_{j_1}}} \exp\left(-\bar{m}_{j_1}F_{{\rm d}_{j_1}}\left(\bar{P}_{j_1,i}r|{ z}\right)\right) \prod\limits_{j_2\in{\cal K}_2}\exp\left(-\pi\lambda_{j_2}\bar{P}_{j_2,i}^2r^2\right)r,\\  r\geq 0, \text{if }i\in{\cal K}_2. 
\end{multline}
\end{subequations}
\end{lemma}
\begin{IEEEproof}
See Appendix~\ref{app::serving::distance::distribtuion}.
\end{IEEEproof}
\subsection{Coverage Probability}\label{sub::sec::coverage::probability}
{Before deriving the main results on coverage probability, we first introduce PGFL and sum-product functional of a point process  which will be appearing repeatedly into the coverage analysis.} 
\begin{ndef}[PGFL]\label{def::pgfl} PGFL of a point process $\Phi$ is defined as:
$\nbbE\left[\prod\limits_{{\bf x}\in\Phi}\mu({\bf x})\right]$, where $\mu({\bf x}):\R^2\to[0,1]$ is measurable. 
\end{ndef}
\begin{lemma}\label{lemm::pgfl::PPP}
	When $\Phi(\lambda({\bf x}))$ is a PPP, the PGFL is given as~\cite{chiu2013stochastic}:
	\begin{align}\label{eq::pgfl::PPP}
		\nbbE\left[\prod\limits_{{\bf x}\in\Phi}\mu({\bf x})\right]  = \exp\left(-\int\limits_{\R^2}\lambda({\bf x})(1-\mu({\bf x})){\rm d}{\bf x}\right).
	\end{align}
\end{lemma}
When  $\mu({\bf x})=\mu(x)$ and $\Phi(\lambda)$ is homogeneous, then \eqref{eq::pgfl::PPP} becomes:
\begin{equation}\label{eq::pgfl::PPP::1d}
	\nbbE\left[\prod\limits_{{\bf y}\in\Phi}\mu({y})\right]  = \exp\left(-2\pi\lambda\int\limits_{0}^{\infty}(1-\mu({ y}))y{\rm d}{y}\right).
\end{equation}
\begin{ndef}[Sum-product functional]\label{def::sum::profuct} Sum-product functional of a point process $\Phi$ is defined in this paper as $\nbbE\left[\sum\limits_{{\bf x}\in{\Phi}}\nu({\bf x})\prod\limits_{{\bf y}\in\Phi}\mu({\bf y})\right]$, 
where $\nu({\bf x}):\nbbR^2\to\R^+$ and $\mu({\bf y}):\nbbR^2\to[0,1]$ are measurable. 
\end{ndef}
{For a PPP, the sum-product functional is given by the following Lemma.} 
\begin{lemma}\label{lemm::sum::product::PPP}
	When $\Phi(\lambda({\bf x}))$ is a PPP, the sum-product functional is given as~\cite{HaeInterferenceFucntionals,saha20173gpp}
	\begin{align}\label{eq::sum::product::PPP}
		\nbbE\left[\sum\limits_{{\bf x}\in{\Phi}}\nu({\bf x})\prod\limits_{{\bf y}\in\Phi}\mu({\bf y})\right]  = \int\limits_{\R^2}\lambda({\bf x}) \nu({\bf x})\mu({\bf x}){\rm d}{\bf x}\exp\left(-\int\limits_{\R^2}\lambda({\bf y})(1-\mu({\bf y})){\rm d}{\bf y}\right).
	\end{align}
\end{lemma}
When ${\nu}({\bf x})\equiv {\nu}(x)$, $\mu({\bf x})\equiv \mu(x)$, and $\Phi(\lambda)$ is homogeneous,  then \eqref{eq::sum::product::PPP} becomes:
\begin{equation}
\label{eq::sum::product::PPP::1d}
\nbbE\left[\sum\limits_{{x}\in{\Phi}}\nu({ x})\prod\limits_{{\bf y}\in\Phi}\mu({ y})\right]  = 2\pi\lambda\int\limits_{0}^{\infty} \nu({ x})\mu({ x})x{\rm d}{ x}\exp\left(-2\pi\lambda\int\limits_{0}^{\infty}(1-\mu({y}))y{\rm d}{y}\right).
\end{equation}
Before providing the final expression of coverage probability, we provide an important intermediate expression of the conditional $i^{th}$ tier coverage probability given all parent point processes. In fact, a key contribution of this paper, as will be evident in sequel, is to show that this conditional coverage probability can be factored as a product of standard functionals (such as PGFL and sum-product functional) of the parent PPPs. 
\begin{lemma}\label{lemm::conditional::coverage}
The $i^{th}$ tier coverage probability  given $\Phi_{{\rm p}_k}, \forall\ k\in{\cal K}_1$ is given by 
$\nbbP(\sinr({\bf x}^*)>\tau_i,{\cal S}_i|\Phi_{{\rm p}_k},\forall\ k\in{\cal K}_1) = $  
\begin{subequations}\label{eq::conditional::coverage::1}
\begin{multline}\label{eq::conditional::coverage::1a}
\bar{m}_{i}\int\limits_{0}^{\infty} \exp\left(-\frac{\tau_i {\tt N}_0 r^{\alpha}}{P_i}\right)\prod\limits_{j_1\in{\cal K}_1\setminus\{i\}}\prod\limits_{{\bf z}\in\Phi_{{\rm p}_{j_1}}}
{\cal C}_{j_1,i}(r,z)\prod\limits_{j_2\in{\cal K}_2}
\exp
\left(
-\pi r^2\lambda_{j_2}\bar{P}_{j_2,i}^2\rho(\tau_i,\alpha)
\right)\times\\
\left(\sum\limits_{{\bf z}\in\Phi_{{\rm p}_i}}f_{{\rm d}_i}(r|{z})
\prod\limits_{{\bf z}\in\Phi_{{\rm p}_i}}{\cal C}_{i,i}(r,z)
\right){\rm d}r, 
\text{when }i\in{\cal K}_1,
\end{multline}
\begin{multline}\label{eq::conditional::coverage::1b}
2\pi\lambda_{i}\int\limits_{0}^{\infty} \exp\left(-\frac{\tau_i {\tt N}_0 r^{\alpha}}{P_i}\right)\prod\limits_{j_1\in{\cal K}_1}\prod\limits_{{\bf z}\in\Phi_{{\rm p}_{j_1}}}{\cal C}_{j_1,i}(r,z)\prod\limits_{j_2\in{\cal K}_2}
\exp
\left(-\pi r^2\lambda_{j_2}\bar{P}_{j_2,i}^2\rho(\tau_i,\alpha)
\right)
r{\rm d}r, \text{when }i\in{\cal K}_2,
\end{multline}
\end{subequations} where
\begin{equation}\label{eq::calC}
{\cal C}_{j,k}(r,z)=\exp\left(-\bar{m}_{j}\left(1-\int\limits_{\bar{P}_{j,k}r}^{\infty} \frac{ f_{{\rm d}_{j}}(y|{z})}{1+\tau_k \frac{{(\bar{P}_{j,k} r)}^{\alpha}}{y^{\alpha}}}{\rm d}y \right) \right),
\end{equation}
and, $\rho(\tau_i,\alpha) = 1+\tau_{i}^{2/\alpha} \int\limits_{\tau_i^{-2/\alpha}}^{\infty} 
\frac{1}{1+t^{\alpha/2}}{\rm d}t = \frac{2\tau_i}{\alpha-2}{}_{2}{\cal F}_{1}\left[1,1-\frac{2}{\alpha};2-\frac{2}{\alpha};-\tau_i\right]$, where $_{2}{\cal F}_{1}$ is the Gauss hypergeometric function~\cite{jo2012heterogeneous}. 
\end{lemma}
\begin{IEEEproof}
See Appendix~\ref{app::conditional::coverage}.
\end{IEEEproof}
Looking closely at the expressions of  the $i^{th}$ tier coverage  probability given $\Phi_{{\rm p}_k},\forall\ k\in{\cal K}_1$, we find two terms: (i) $\prod_{{\bf z}\in \Phi_{{\rm p}_{j_1}}}{\cal C}_{j_1,i}(r,z)$ in \eqref{eq::conditional::coverage::1a} and \eqref{eq::conditional::coverage::1b}, and (ii) $\sum_{{\bf z}\in {\Phi}_{{\rm p}_{i}}}f_{{\rm d}_i}(r|z)\prod_{{\bf z}\in \Phi_{{\rm p}_{i}}}{\cal C}_{i,i}(r,z)$ in \eqref{eq::conditional::coverage::1a}. These two terms are respectively product and sum-product over all points of $\Phi_{{\rm p}_k}$ which will be substituted by the  PGFL  and sum-product functional of  $\Phi_{{\rm p}_k}$ while decondtioning over   $\Phi_{{\rm p}_k}$. 

\begin{remark}In order to apply PGFL and sum-product functional expressions of PPP given by \eqref{eq::pgfl::PPP::1d} and \eqref{eq::sum::product::PPP::1d}, respectively, we require the condition: $\int\limits_{0}^{\infty}\log(|{\cal C}_{j_1,i}(r,z)|)z{\rm d}z<\infty,\ \forall\ j_1\in{\cal K}_1$. In  \cite[Appendix~B]{miyoshi2018downlink}, it was shown that this condition always holds for the form of ${\cal C}_{j_1,i}(r,z)$ given by \eqref{eq::calC}.
\end{remark}
Until this point, all results were conditioned on $\Phi_{{\rm p}_k},\forall\ k\in{\cal K}_1$. Remember that in Section~\ref{subsec::k-tier-hetnet}, we introduced two types of spatial interaction between the users and BSs. 
Since, by construction,  these two types differ only in $\Phi_{{\rm p}_{\rm q}}$ for some $q\in {\cal K}_1$,  we were able to treat  \caseS~1 and \caseS~2 within the same analytical framework.    
We now present the final expression of $\pc$ for the two types explicitly by deconditioning the conditional coverage probability over $\Phi_{{\rm p}_k},\forall\ k\in{\cal K}_1$  in the following Theorems. 
\begin{theorem}[\caseS~1]\label{theorem::coverage::case::1}The coverage probability is given as:
\begin{align}\label{eq::coverage::probability::sum}
\pc =\sum\limits_{i=1}^K \nbbP(\sinr>\tau_i,{\cal S}_i),
\end{align}
where, the {$i^{th}$ tier coverage probability}, $\nbbP(\sinr>\tau_i,{\cal S}_i)=$
\begin{subequations}\label{eq::cov::case1}
\begin{multline}
2\pi\lambda_{{\rm p}_i}\bar{m}_{i}\int\limits_{0}^{\infty} \exp\left(-\frac{\tau_i {\tt N}_0 r^{\alpha}}{P_i}\right)\prod\limits_{j_1\in{\cal K}_1}\exp\left(-2\pi\lambda_{{\rm p}_{j_1}}\int\limits_0^{\infty}\left(1-
{\cal C}_{j_1,i}(r,z)\right)z{\rm d}z\right)\times\\
\exp\left(
-\pi r^2\sum\limits_{j_2\in{\cal K}_2}\lambda_{j_2}\bar{P}_{j_2,i}^2\rho(\tau_i,\alpha)
\right)\int_{0}^{\infty}f_{{\rm d}_i}(r|{z})
{\cal C}_{i,i}(r,z){z}{\rm d}z\:{\rm d}r,\text{ for }i\in{\cal K}_1,
\end{multline}
\begin{multline}\label{eq::coverage::theorem1::case2}
2\pi{\lambda}_{i}
\int\limits_{0}^{\infty} \exp\left(-\frac{\tau_i {\tt N}_0 r^{\alpha}}{P_i}\right)\prod\limits_{j_1\in{\cal K}_1}\exp\left(-2\pi\lambda_{{\rm p}_{j_1}}\int\limits_0^{\infty}\left(1-
{\cal C}_{j_1,i}(r,z)\right)z{\rm d}z\right)\times
\\\exp
\left(
-\pi r^2\sum\limits_{j_2\in{\cal K}_2}\lambda_{j_2}\bar{P}_{j_2,i}^2\rho(\tau_i,\alpha)
\right)r{\rm d}r, \text{ for }i\in{\cal K}_2.
\end{multline}
\end{subequations}
\end{theorem}
\begin{IEEEproof}When $i\in{\cal K}_1$, we get from \eqref{eq::conditional::coverage::1a},  $\nbbP(\sinr>\tau_i,{\cal S}_i) = \nbbE\left[\nbbP(\sinr>\tau_i,{\cal S}_i|{\Phi_{{\rm p}_k},\forall\ k\in{\cal K}_1}) \right] = $
\begin{multline*}
 \bar{m}_{i}\int\limits_{0}^{\infty} \exp\left(-\frac{\tau_i {\tt N}_0 r^{\alpha}}{P_i}\right)\prod\limits_{j_1\in{\cal K}_1\setminus\{i\}}\nbbE_{\Phi_{{\rm p}_{j_1}}}\left[\prod\limits_{{\bf z}\in\Phi_{{\rm p}_{j_1}}}
{\cal C}_{j_1,i}(r,z)\right]
\exp\left(-\pi r^2\sum\limits_{j_2\in{\cal K}_2}\lambda_{j_2}\bar{P}_{j_2,i}^2\rho(\tau_i,\alpha)\right)\times\\\nbbE_{\Phi_{{\rm p}_i}}\left[\sum\limits_{{\bf z}\in\Phi_{{\rm p}_i}}f_{{\rm d}_i}(r|{z})
\prod\limits_{{\bf z}\in\Phi_{{\rm p}_i}}{\cal C}_{i,i}(r,z)
\right]{\rm d}r.
\end{multline*}
This step is enabled by the assumption that $\Phi_{j}$-s are independent $\forall\ j\in {\cal K}$.  
The final expression is obtained by substituting the PGFL of $\Phi_{{\rm p}_k}$ for $k\in{\cal K}_1\setminus\{i\}$ from \eqref{eq::pgfl::PPP::1d} and the sum-product functional of $\Phi_{{\rm p}_i}$ from  \eqref{eq::sum::product::PPP::1d}.  The final expression follows from some algebraic simplifications.   Now, 
for $i\in{\cal K}_2$, from \eqref{eq::conditional::coverage::1b}, we get
\begin{multline*}
\nbbP(\sinr>\tau_i,{\cal S}_i) = 
2\pi\lambda_{i}\int\limits_{0}^{\infty} \exp\left(-\frac{\tau_i {\tt N}_0 r^{\alpha}}{P_i}\right)\prod\limits_{j_1\in{\cal K}_1}\nbbE_{\Phi_{{\rm p}_{j_1}}}\left[\prod\limits_{{\bf z}\in\Phi_{{\rm p}_{j_1}}}
{\cal C}_{j_1,i}(r,z)\right]\times\\
\exp
\left(-\pi r^2\sum\limits_{j_2\in{\cal K}_2}\lambda_{j_2}\bar{P}_{j_2,i}^2\rho(\tau_i,\alpha)
\right)
r{\rm d}r.
\end{multline*}
We then use the PGFL of $\Phi_{{\rm p}_{j_1}}$ from \eqref{eq::pgfl::PPP::1d} to obtain the final expression.
\end{IEEEproof}
\begin{theorem}[\caseS~2]\label{theorem::coverage::case::2}
The coverage probability $\pc$ can be written as \eqref{eq::coverage::probability::sum}, where  the $i^{th}$ tier  coverage probability is: $\nbbP(\sinr>\tau_i,{\cal S}_i)=$
\begin{subequations}\label{eq::cov::case2}
\begin{multline}
2\pi\lambda_{{\rm p}_i}\bar{m}_{i}\int\limits_{0}^{\infty} \exp\left(-\frac{\tau_i {\tt N}_0 r^{\alpha}}{P_i}\right)\prod\limits_{j_1\in{\cal K}_1}\exp\left(-2\pi\lambda_{{\rm p}_{j_1}}\int\limits_0^{\infty}\left(1-
{\cal C}_{j_1,i}(r,z)\right)z{\rm d}z\right)
\int\limits_{0}^{\infty}{\cal C}_{q,i}(r,z_0)
f_{{\rm d}_{\rm u}}(z_0|0){\rm d}z_0\times\\\exp\left(
-\pi r^2\sum\limits_{j_2\in{\cal K}_2}\lambda_{j_2}\bar{P}_{j_2,i}^2\rho(\tau_i,\alpha)
\right)\int\limits_{0}^{\infty}f_{{\rm d}_i}(r|{z})
{\cal C}_{i,i}(r,z){z}{\rm d}z\:{\rm d}r,\text{ for }i\in{\cal K}_1\setminus \{q\},
\end{multline}
\begin{multline}
\bar{m}_{q}\int\limits_{0}^{\infty} \exp\left(-\frac{\tau_i {\tt N}_0 r^{\alpha}}{P_q}\right)\prod\limits_{j_1\in{\cal K}_1}\exp\left(-2\pi\lambda_{{\rm p}_{j_1}}\int\limits_0^{\infty}\left(1-
{\cal C}_{j_1,i}(r,z)\right)z{\rm d}z\right)
\exp\left(
-\pi r^2\sum\limits_{j_2\in{\cal K}_2}\lambda_{j_2}\bar{P}_{j_2,q}^2\rho(\tau_q,\alpha)
\right)\times\\
\bigg(\int\limits_0^{\infty}f_{{\rm d}_q}(r|z_0){\cal C}_{q,q}(r,z_0)f_{{\rm d}_{\rm u}}(z_0|0){\rm d}{z_0}
+2\pi\lambda_{{\rm p}_q}\int\limits_{0}^{\infty}f_{{\rm d}_q}(r|z){\cal C}_{q,q}(r,z)z{\rm d}z\times\\\int\limits_{0}^{\infty}{\cal C}_{q,q}(r,z_0)f_{{\rm d}_{\rm u}}({z_0}|0){\rm d}z_0\bigg){\rm d}r,\text{ for }i=q,
\end{multline}
\begin{multline}
2\pi\lambda_{i}\int\limits_{0}^{\infty} \exp\left(-\frac{\tau_i {\tt N}_0 r^{\alpha}}{P_i}\right)\prod\limits_{j_1\in{\cal K}_1}\exp\left(-2\pi\lambda_{{\rm p}_{j_1}}\int\limits_0^{\infty}\left(1-
{\cal C}_{j_1,i}(r,z) \right)z{\rm d}z\right)\times\\
\int\limits_{0}^{\infty}{\cal C}_{q,i}(r,z_0)
f_{{\rm d}_{\rm u}}(z_0|0){\rm d}z_0\exp\left(
-\pi r^2\sum\limits_{j_2\in{\cal K}_2}\lambda_{j_2}\bar{P}_{j_2,i}^2\rho(\tau_i,\alpha)\right)r{\rm d}r,\text{ for }i\in{\cal K}_2.
\end{multline}
\end{subequations}
Here $f_{{\rm d}_{\rm u}}(z_0|0)$ denotes the distance distribution of a point of $\Phi_{\rm u}$ from its cluster center which resides at origin.  
\end{theorem}
\begin{IEEEproof}
See Appendix~\ref{app::theorem::coverage::case::2}. 
\end{IEEEproof}
We conclude this discussion with the following remark. 
\begin{remark}\label{rem::remark::after::coverage}
The analytical framework developed in this Section provisions to model the $k^{th}$ BS tier, where $k\in {\cal K}_1$ as any arbitrary PCP $\Phi(\lambda_{{\rm p}_k},g_k,\bar{m}_k)$. By looking into the expressions of $\pc$ in Theorems~\ref{theorem::coverage::case::1} and \ref{theorem::coverage::case::2},  it is apparent that given $K_1$ PCPs with arbitrary distributions, i.e.,  $\{\Phi_k = \Phi(\lambda_{{\rm p}_k},g_k,\bar{m}_k), k\in{\cal K}_1\}$, the only non-trivial step is to find the conditional distance distributions $\{f_{{\rm d}_k}({x}|z)\}$ which need to be simply plugged into the expressions of $\nbbP(\sinr>\tau_i,{\cal S}_i)$ in \eqref{eq::cov::case1}-\eqref{eq::cov::case2}. 
\end{remark}
The unified model discussed in this paper reduces to the conventional PPP-based HetNet model with no spatial coupling between users and BS locations by setting ${\cal K}_1 = \emptyset$. For  this scenario, \Case~2 becomes irrelevant and the coverage probability is given by the following Corollary.
\begin{cor}\label{cor::cov::no::corr}
Setting ${\cal K}_1 = \emptyset$, coverage probability for \Case~1 is given by:
\begin{align}\label{eq::cov::no::corr}
\pc = \sum\limits_{i\in {\cal K}_2} 2\pi\lambda_{i}\int\limits_{0}^{\infty}\exp\left(-\frac{\tau_i{\tt N}_0r^{\alpha}}{P_i}\right)\exp\left(-\pi r^2\sum\limits_{j\in{\cal K}_2}\lambda_{j_2}\bar{P}_{j,i}^2\rho(\tau_i,\alpha)\right)r{\rm d}r,
\end{align}
and, for interference-limited networks (${\tt N}_0 =  0 $),
\begin{equation}\label{eq::cov::no::corr::nonoise}
\pc = \frac{\sum\limits_{i\in{\cal K}_2}\lambda_{i}P_i^{\frac{2}{\alpha}}}{\sum\limits_{j\in{\cal K}_2}\lambda_j P_j^{\frac{2}{\alpha}}\rho(\tau_i,\alpha)}.
\end{equation}
\end{cor} 
\begin{IEEEproof} This result can be obtained directly from \eqref{eq::coverage::theorem1::case2} by setting ${\cal K}_1=\emptyset$. See \cite{jo2012heterogeneous,andrews2016primer} for the intermediate steps between \eqref{eq::cov::no::corr} and \eqref{eq::cov::no::corr::nonoise}. 
\end{IEEEproof}
\subsection{Special Cases: TCP and MCP}
 For the purpose of numerical evaluation of coverage probability, we assume that $\Phi_{\rm k}$ is either a TCP or an MCP which are defined as follows. 
\begin{ndef}[Thomas Cluster Process]\label{def::TCP}
 A PCP $\Phi_k\ (\lambda_{{\rm p}_k},g_k,\bar{m}_k)$ is called a  TCP if the distribution of the offspring points in $\ncalB_k^{\bf z}$ is Gaussian around the cluster center at the  origin, i.e. for all ${\bf s}\in\ncalB_k^{\bf z}$,  if ${\bf s}=(\|{\bf s}\|,\arg({\bf s}))\equiv(s,\theta_s)\in\ncalB_k^{\bf z}$ denotes a point of the offspring point process ${\ncalB_k^{\bf z}}$ with cluster center at origin, 
\begin{align}\label{eq::density_thomas_definition}
g_k({\bf s}) =g_k( {s},\theta_s)=  \frac{s}{\sigma_k^2}\exp\left(-\frac{s^2}{2\sigma_k^2}\right)\frac{1}{2\pi},\ s\geq 0,  0<\theta_s\leq 2\pi.
\end{align}  
\end{ndef}
\begin{ndef}[\matern Cluster Process] A PCP  $\Phi_k\ (\lambda_{{\rm p}_k},g_k,\bar{m}_k)$ is called an MCP if the distribution of the offspring points in $\ncalB_k^{\bf z}$ is uniform within a disc of radius $r_{{\rm d}_k}$ around the origin denoted by $b({\bf 0},r_{{\rm d}_k})$. 
Hence, 
\begin{align}\label{eq::density_matern_definition}
g_k({\bf s}) =g_k(s,\theta_s) = \frac{2s}{r_{{\rm d}_k}^2}{\times\frac{1}{2\pi}}, \qquad 0\leq s\leq r_{{\rm d}_k},0< \theta_s\leq 2\pi.
\end{align}  
\end{ndef}
  We now provide the conditional distance distributions of TCP and MCP. 
  When $\Phi_k$ is a TCP, 
given that ${\bf z}$ is the cluster center of $
{\bf x}$, i.e., ${\bf x}\in{\bf z}+{\cal B}_k^{\bf z}$, we can write the conditional PDF of $x$
 as \cite{AfshDhi2015MehrnazD2D1}:
 \begin{align}\label{eq::marginal::dist::tcp}
f_{{\rm d}_k}(x|z)
= \frac{x}{ \sigma_k^2} \exp\left(-\frac{x^2+z^2}{2 \sigma_k^2}\right) I_0\left(\frac{x z}{\sigma_k^2}\right), \qquad x,z\geq 0,
\end{align} where $I_0(\cdot)$ is the modified Bessel function of the first kind with order zero, and,
  \begin{align}\label{eq::marginal::dist::tcp::z0}
f_{{\rm d}_k}(x|0)
= \frac{x}{ \sigma_k^2} \exp\left(-\frac{x^2}{2 \sigma_k^2}\right), \qquad x\geq 0.
\end{align}
When $\Phi_k$ is an MCP, $f_{{\rm d}_k}({x|z})=
\chi_k^{(\ell)}(x,z),\ (\ell=1,2)$,
 where 
\begin{subequations}\label{eq::chi_definition}
\begin{alignat}{4}
&\chi_k^{(1)}(x,z) = \frac{2 x}{r_{{\rm d}_k}^2},& 0\leq x
\leq r_{{\rm d}_k}-z, 0\leq z\leq r_{{\rm d}_k},\\
&\chi_k^{(2)}(x,z) = \frac{2 x}{\pi r_{{\rm d}_k}^2}\cos^{-1}\left(\frac{x^2+z^2-r_{{\rm d}_k}^2}{2xz}\right), &|r_{{\rm d}_k}-z|<x\leq r_{{\rm d}_k}+z.
\end{alignat}
\end{subequations}
\section{Results and Discussions}
\begin{figure}
          \centering
          \subfigure[$\Phi_1$ is a TCP.]{
              \includegraphics[width=.45\linewidth]{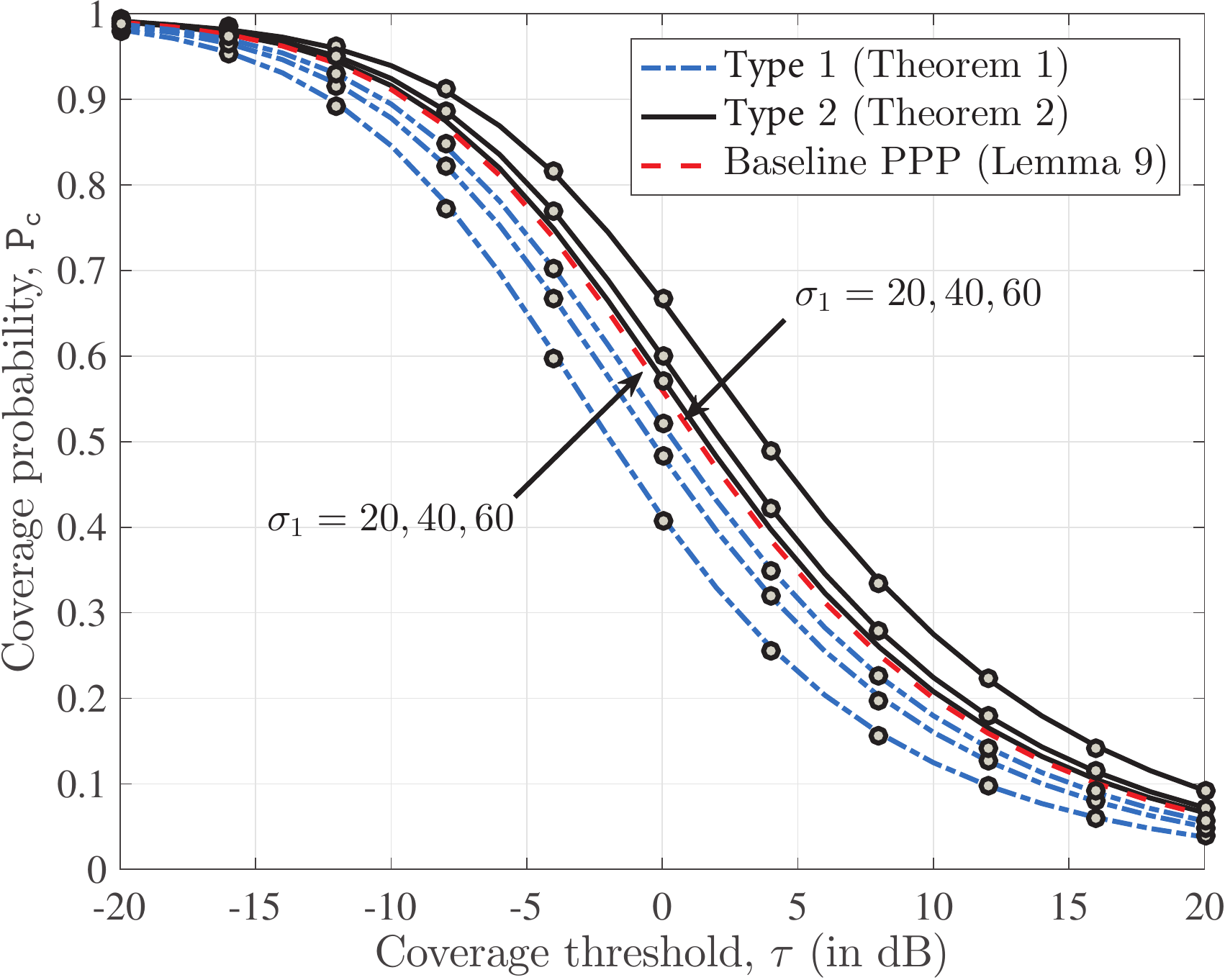}
             \label{fig::validation::thomas}}
              \subfigure[$\Phi_1$ is an MCP.]{
               \label{fig::validation::matern}
              \includegraphics[width=.45\linewidth]{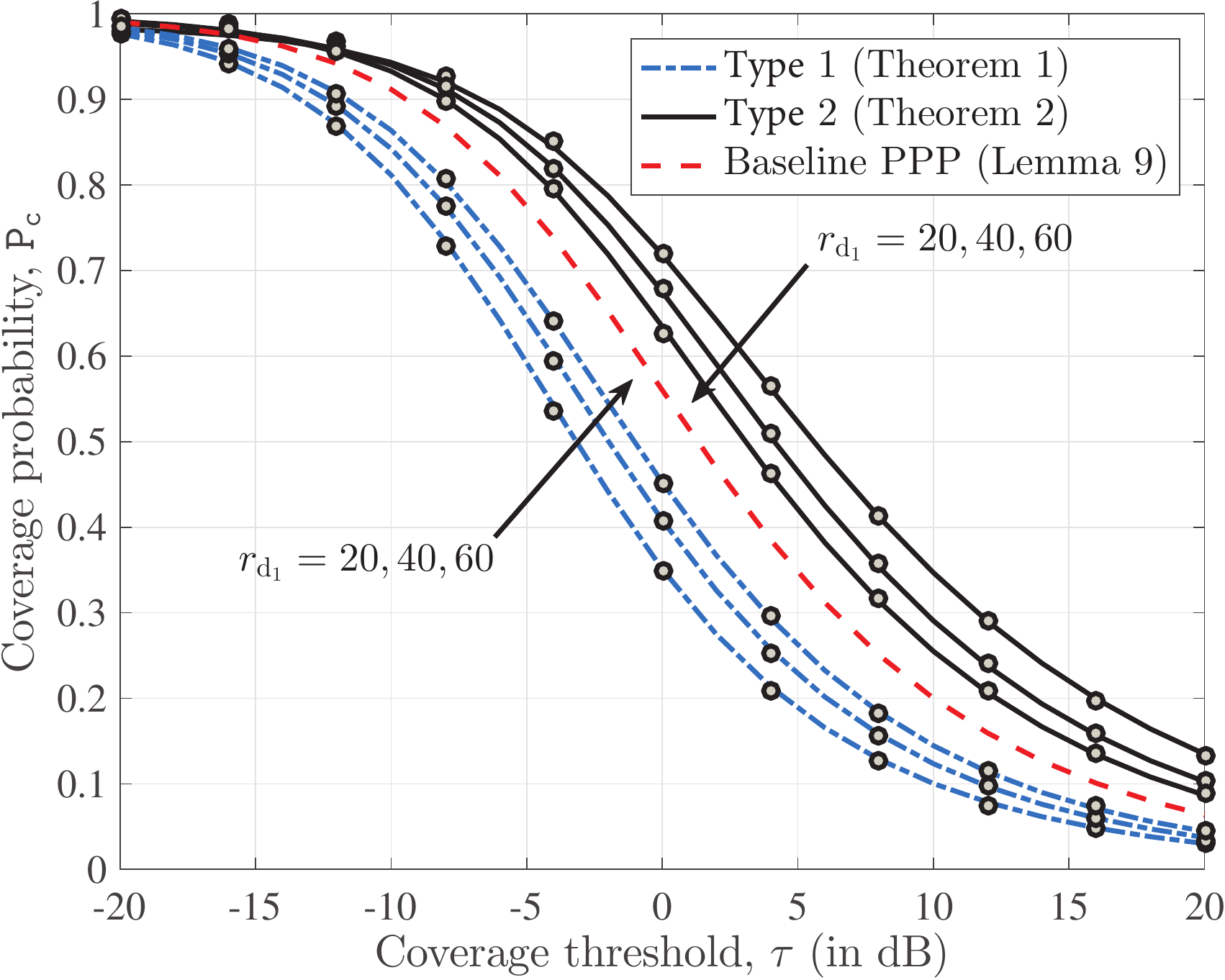}
              }
              \caption{Coverage probability as a function of $\sir$ threshold ($\alpha=4$, $P_2 = 10^3P_1$, $\lambda_2 = 10^{-6}\ \text{m}^{-2}$,  $\lambda_{{\rm p}_1} = 10^{-4}\ \text{m}^{-2}$, and  $\bar{m}=10$ ).}
             \label{fig::validation}
          \end{figure}

We now numerically evaluate the expressions of $\pc$ derived in Theorems~\ref{theorem::coverage::case::1} and \ref{theorem::coverage::case::2}. For the numerical evaluation,   we choose a two tier network ($K=2$) with one tier of  sparsely deployed  MBSs and another tier of densely deployed  SBSs. 
 The  SBSs are  distributed as a PCP and the MBSs are distributed as a PPP, i.e., ${\cal K}_1 = \{1\}$ and ${\cal K}_2 =\{ 2\}$. We assume that the network is interference limited (${\tt N}_0 = 0$) and the downlink transmit powers of each BS in $\Phi_1$ and $\Phi_2$ are set such that   $P_2/P_1=10^3$.   Also,  $\Phi_1$ is assumed to be denser than $\Phi_2$, i.e., $\bar{m}_1\lambda_{{\rm p}_1}>\lambda_2$. We also assume $\alpha = 4$ and $\tau_1=\tau_2=\tau$. We first verify the analytical results presented in Theorems~\ref{theorem::coverage::case::1} and \ref{theorem::coverage::case::2} with Monte Carlo simulations of the network, for which we set $\bar{m}=10, \lambda_{{\rm p}_1}= 100\ \text{km}^{-2}$ and $\lambda_{2}= 1\ \text{km}^{-2}$.  The values of $\pc$ for different values of $\tau$ from simulation and analysis are plotted in Fig.~\ref{fig::validation}, where  Fig.~\ref{fig::validation::thomas}  and \ref{fig::validation::matern} corresponds to $\Phi_1$ being TCP and MCP, respectively. {The perfect match between simulation  and analytical results, indicated by small circles and curves, respectively,  verifies the accuracy of our analysis.} 
 \subsection{Variation of Cluster Size}
In Fig.~\ref{fig::validation}, $\pc$ is plotted for different cluster sizes of $\Phi_1$. The cluster size precisely refers to  $\sigma_1$ when $\Phi_1$ is a TCP (in Fig.~\ref{fig::validation::thomas}) and ${r_{{\rm d}_1}}$ when $\Phi_1$ is an MCP (in Fig.~\ref{fig::validation::matern}). We observe that the cluster size has a conflicting effect on $\pc$: for \Case~1, $\pc$ increases with cluster size and for \Case~2, $\pc$ decreases with cluster size. This can be explained as follows. In \Case~2, due to the spatial coupling between $\Phi_1$ and $\Phi_{\rm u}$, the candidate serving BS of $\Phi_1$ is more likely to belong to the representative cluster (i.e., the BS  cluster with the  cluster center of the typical user) and as cluster size increases, this candidate serving BS moves farther away from the user {on average}. On the other hand, for \Case~1, as cluster size increases, the BSs of $\Phi_1$ {on average} lie closer to the typical user. This phenomenon is the consequence of  the spatial coupling between BSs and users and is also observed for the  $\max$-$\sinr$ based association strategy      
 in the similar setup~\cite{saha20173gpp}. For both types, as cluster size increases, $\pc$ converges to the coverage probability for a two tier network with both BS tiers being modeled as PPP, more precisely, ${\cal K}_1 =\emptyset$, ${\cal K}_2 = \{1,2\}$, with intensities  $\lambda_1 = \bar{m}_1\lambda_{{\rm p}_1}$ and $\lambda_2$, respectively.  The reason for this convergence is the fact that  as cluster size  tends to infinity,  the limiting distribution of a PCP is a PPP~\cite{chiu2013stochastic}.  
   Since the trends of $\pc$ are very similar for $\Phi_1$ being TCP and MCP, we set  $\Phi_1$ as TCP for the rest of the discussion.  

 Another interesting observation from  Fig.~\ref{fig::validation}  is  that the variation of $\pc$ with cluster size is not as prominent for \Case~2 as it is for \Case~1. For further investigation, we focus on the scenario where $\Phi_1$ is a TCP,  fix $\tau = 0\ \text{dB}$ and plot the variation of $\pc$ with $\sigma_1$ in Fig.~\ref{fig::varysigma}. We observe that the $\pc$ versus $\sigma_1$ curves for \Case~2 are almost flat. The reason can be explained as follows. For \Case~2, as cluster size is increased, the nearest BS of $\Phi_1$ belonging to the representative cluster lies farther away while the nearest BS of $\Phi_1$ which does not belong to the representative cluster comes closer.  Also, as cluster size increases, the intra-cluster interference, i.e., the aggregate interference from the BSs of the representative cluster decreases while the inter-cluster interference, i.e., the aggregate interference from the BSs of $\Phi_1$ except the representative cluster increases. Due to these conflicting effects, the cluster size variation do not impact $\pc$ strongly for \Case~2. On the other hand, for \Case~1 these conflicts are not present because $\Phi_{\rm u}$ and $\Phi_1$ are not coupled and hence there is no representative BS cluster. 
 
 \begin{minipage}{\linewidth}
      \centering
      \begin{minipage}{0.45\linewidth}
 \begin{figure}[H]
          \centering
              \includegraphics[width=.9\linewidth]{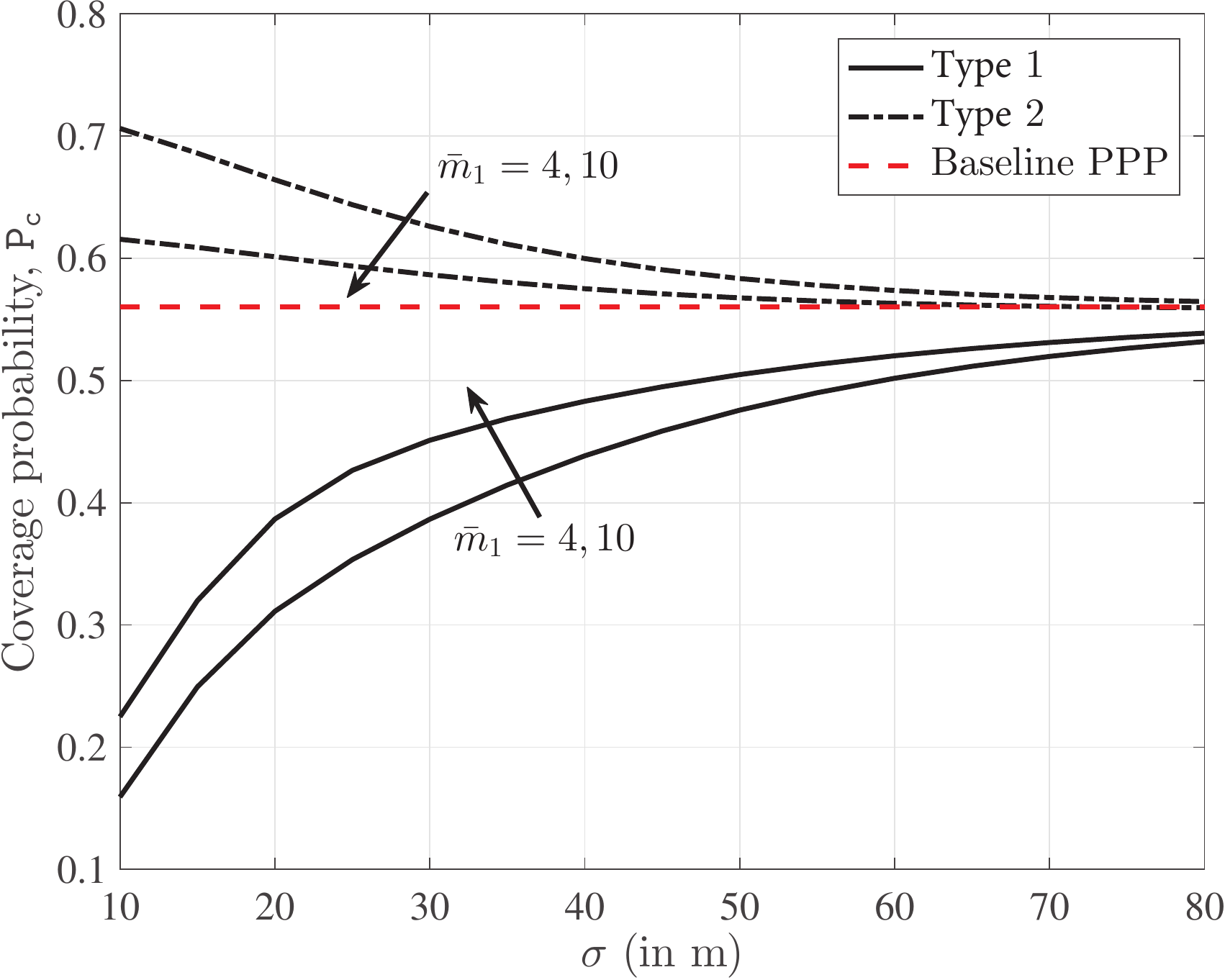}
              \caption{Coverage probability as a function of $\sigma_1$ ($\alpha=4$, $P_2 = 10^3P_1$, $\lambda_{{\rm p}_1} = 10^{-4}\ \text{m}^{-2}$, $\lambda_2 = 10^{-6}\ \text{m}^{-2}$, and $\tau = 0\ \text{dB}$).}
             \label{fig::varysigma}
\end{figure}       
      \end{minipage}
      \hspace{0.05\linewidth}
      \begin{minipage}{0.45\linewidth}
    \begin{figure}[H]
          \centering
              \includegraphics[width=.9\linewidth]{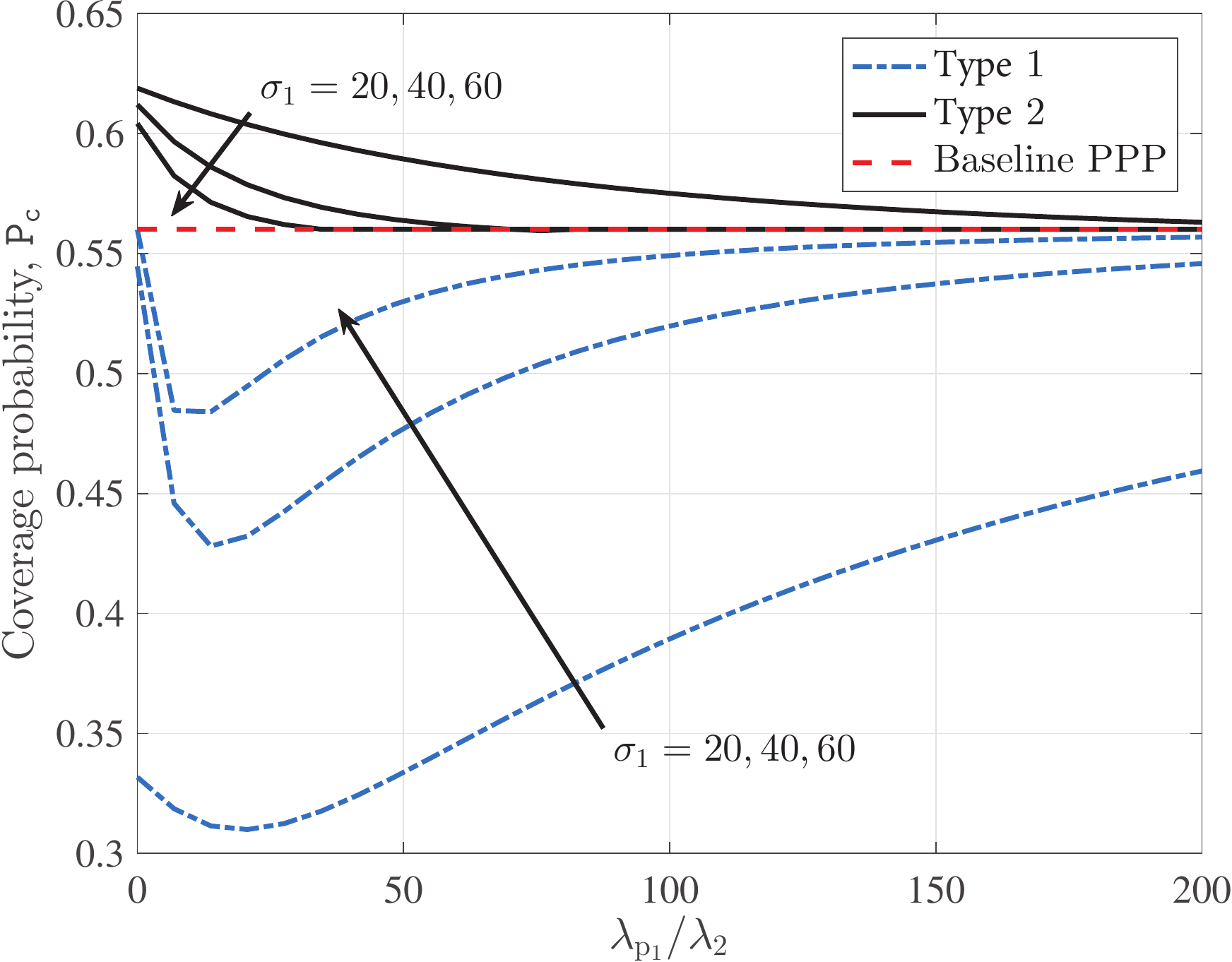}
         \caption{Coverage probability as a function of $\lambda_{{\rm p}_1}/\lambda_2$ ($\alpha=4$, $P_2 = 10^3P_1$, $\tau = 0\ \text{dB}$, and  $\bar{m}=10$).\newline}
             \label{fig::varylam}
 \end{figure}         

      \end{minipage}
  \end{minipage}   
          
\subsection{Variation of BS Intensity}
We now vary the intensity of the parent PPP ($\lambda_{{\rm p}_1}$) keeping $\lambda_2$ constant and plot the coverage probability in Fig.~\ref{fig::varylam}. For an interference-limited HetNet with the  same $\sir$ threshold for all tiers, it is a well-known result that if the BSs are modeled as PPPs, $\pc$ is independent of the BS intensity. This can be readily verified by putting $\tau_i=\tau$ $\forall\ i$ in \eqref{eq::cov::no::corr::nonoise} which yields  $\pc = {1}/{\rho(\tau,\alpha)}$. However, once the spatial  distribution of  BSs is changed to  PCP, we see that $\pc$ is rather strongly dependent on the BS intensity under similar set of assumptions.  From Fig.~\ref{fig::varylam}, we also observe that for large values  of  $\lambda_{{\rm p}_1}$, $\pc$ converges to ${1}/{\rho(\tau,\alpha)}$, i.e., the coverage when the BSs are modeled as PPPs.
\begin{remark}
Although it is difficult to visualize the variation of $\pc$ in the parameter space, which  for the two tier network under consideration is $\lambda_{{\rm p}_1}\times \lambda_2\times\bar{m}_1\times\sigma_1$, using the results in \cite{afshang2018equi}, it is possible to find the trajectories (known as equi-coverage contours) in the parameter space along which $\pc$ remains constant. The family of equicoverage contours are generated by $(\lambda_{{\rm p}_1}/l^2,\lambda_2/l^2,\sigma_1 l)$ where $l>0$ is a scalar. When $\Phi_1$  is an MCP, a similar result can be obtained by replacing $\sigma_1$ by $r_{{\rm d}_1}$.  
\end{remark}
\section{Conclusion}
Although it is more realistic to model BS and user locations of a HetNet as  PCPs, the exact characterization of coverage probability for the $\max$-power based association strategy for PCP distributed BSs has remained an open problem for a while. The main contribution of this paper is a concrete resolution of this problem. In particular, we developed an analytical framework for evaluating the coverage probability of a typical user in a $K$-tier HetNet model where the locations of a fraction of BS tiers are modeled as PCP and the rest are modeled as PPP. To be consistent with  3GPP  HetNet models, 
we also assumed that the user distribution is either independent of or spatially coupled to the BS distribution. 
This work, along with our previous work~\cite{saha20173gpp} (focused  on $\max$-$\sinr$ based association), provides a complete characterization of downlink coverage probability in the unified HetNet model, which is an important generalization of the well-known baseline PPP-based model. 

This work has numerous extensions. An immediate extension is to study other key performance  metrics such as rate coverage probability and percentile rates which will take into account the distributions of load or the number of users connected to the BSs. From stochastic geometry perspective, this will necessitate  characterizing the  Voronoi cells generated by PCP-distributed BSs whose properties are not well-studied in the literature. From  a modeling perspective, another extension is to incorporate more realistic channel models including shadowing,  blocking and general pathloss, e.g. in the context of millimeter wave (mm-wave) communication. Especially for  mm-wave integrated access and backhaul (IAB) in 5G~\cite{saha2018IAB}, the PCP is a natural candidate for modeling  the locations of the  SBSs forming clusters around the MBS that  provides wireless backhaul over mmWave links.  More generally, the framework developed in this paper will lay the foundation to  
 the performance  analysis of such networks where the spatial coupling between the BS and user locations cannot be ignored.
\appendix
\subsection{Proof of Lemma~\ref{lemm::association::probability::PCP}}
\label{app::association::probability::PCP}
When $i\in{\cal K}_1$,
\begin{align*}
&\nbbP({\cal S}_i|\Phi_{{\rm p}_k}, \forall\ k\in{\cal K}_1)=\nbbP\left(\bigcap\limits_{j\in{\cal K}\setminus\{i\}}P_i \|\widetilde{{\bf x}}_i\|^{-\alpha}>P_j \|\widetilde{{\bf x}}_{j}\|^{-\alpha}\bigg| \Phi_{{\rm p}_k}, \forall\ k\in{\cal K}_1\right)\\
&=
\prod\limits_{j_1\in{\cal K}_1\setminus\{i\}}\nbbP\left(\|{\widetilde{\bf x}}_{j_1}\|>\left(\frac{P_{j_1}}{P_i}\right)^{1/\alpha}\|{\widetilde{\bf x}}_i\|\bigg|\Phi_{{\rm p}_i},\Phi_{{\rm p}_{j_1}}\right)\prod\limits_{j_2\in{\cal K}_2}\nbbP\left(\|{\widetilde{\bf x}}_{j_2}\|>\left(\frac{P_{j_2}}{P_i}\right)^{1/\alpha}\|{\widetilde{\bf x}}_i\|\bigg|\Phi_{{\rm p}_i}\right) \\
&\stackrel{(a)}{=} \int_{0}^{\infty}\prod\limits_{j_1\in{\cal K}_1\setminus\{i\}}\bar{F}_{c_{j_1}}\left(\left(\frac{P_{j_1}}{P_i}\right)^{1/\alpha}r\bigg|\Phi_{{\rm p}_{j_1}}\right)
\prod\limits_{j_2\in{\cal K}_2}\bar{F}_{c_{j_2}}\left(\left(\frac{P_{j_2}}{P_i}\right)^{1/\alpha}r\right)
f_{c_i}(r|\Phi_{{\rm p}_{i}}){\rm d}{r} \\
&\stackrel{(b)}{=}\int_0^{\infty}\prod\limits_{{j_1}\in{\cal K}_1\setminus\{i\}}\prod_{{\bf z}\in\Phi_{{\rm p}_{j_1}}} \exp\left(-\bar{m}_{j_1}F_{{\rm d}_{j_1}}\left(\bigg(\frac{P_{j_1}}{P_i}\bigg)^{1/\alpha}r\bigg|{z}\right)\right)
\prod\limits_{{j_2}\in{\cal K}_2}\exp\left(-\pi\lambda_{j_2}\bigg(\frac{P_{j_2}}{P_i}\bigg)^{2/\alpha}r^2\right)\times
\\&
\bar{m}_i\sum\limits_{{\bf z}\in\Phi_{{\rm p}_i}}f_{{\rm d}_i}(r|{ z}) \prod\limits_{{\bf z}\in\Phi_{{\rm p}_i}}\exp\left(-\bar{m}_iF_{{\rm d}_i}(r|{ z})\right){\rm d}r\\
&=\bar{m}_i\int_{0}^{\infty}\sum\limits_{{\bf z}\in\Phi_{{\rm p}_i}}f_{{\rm d}_i}(r|{ z}) \prod_{j_1\in{\cal K}_1}\prod_{{\bf z}\in\Phi_{{\rm p}_{j_1}}} \exp\left(-\bar{m}_{j_1}F_{{\rm d}_{j_1}}\left(\bigg(\frac{P_{j_1}}{P_i}\bigg)^{1/\alpha}r\bigg|{z}\right)\right)  \exp\left(-\pi\sum_{j_2\in{\cal K}_2}\lambda_{j_2}\bigg(\frac{P_{j_2}}{P_i}\bigg)^{2/\alpha}r^2\right){\rm d}r. 
\end{align*}
Here $(a)$ follows from the fact that $\Phi_{i}$-s are independent,  $(b)$ follows from Lemma~\ref{lemm::contact::distance::PCP}, and  $\bar{F}_{{\rm c}_{j}}(\cdot)$ denotes the complementary CDF (CCDF) of $\|\widetilde{\bf x}_j\|$. When     $i\in{\cal K}_2$, 
\begin{align*}
&\nbbP({\cal S}_i|\Phi_{{\rm p}_k}, \forall\ k\in{\cal K}_1)\\&=
\prod\limits_{j_1\in{\cal K}_1}\nbbP\left(\|{\widetilde{\bf x}}_{j_1}\|>\left(\frac{P_{j_1}}{P_i}\right)^{1/\alpha}\|{\widetilde{\bf x}}_i\|\bigg|\Phi_{{\rm p}_{j_1}}\right)\prod\limits_{j_2\in{\cal K}_2\setminus\{i\}}\nbbP\left(\|{\widetilde{\bf x}}_{j_2}\|>\left(\frac{P_{j_2}}{P_i}\right)^{1/\alpha}\|{\widetilde{\bf x}}_i\|\right) \\
&= \int_{0}^{\infty}\prod\limits_{j_1\in{\cal K}_1}\bar{F}_{c_{j_1}}\left(\left(\frac{P_{j_1}}{P_i}\right)^{1/\alpha}r\bigg|\Phi_{{\rm p}_{j_1}}\right)
\prod\limits_{j_2\in{\cal K}_2\setminus\{i\}}\bar{F}_{c_{j_2}}\left(\left(\frac{P_{j_1}}{P_i}\right)^{1/\alpha}r\right)
f_{c_i}(r){\rm d}{r} \\
&=2\pi\lambda_i\int_{0}^{\infty}\prod_{j_1\in{\cal K}_1}\prod_{{\bf z}\in\Phi_{{\rm p}_{j_1}}} \exp\left(-\bar{m}_{j_1}F_{{\rm d}_{j_1}}\left(\bigg(\frac{P_{j_1}}{P_i}\bigg)^{1/\alpha}r\bigg|{z}\right)\right)  \exp\left(-\pi\sum_{j_2\in{\cal K}_2}\lambda_{j_2}\bigg(\frac{P_{j_2}}{P_i}\bigg)^{2/\alpha}r^2\right)r{\rm d}r, 
\end{align*}
where the last step follows from Lemma~\ref{lemm::contact::distance::PCP}.
\subsection{Proof of Lemma~\ref{lemm::serving::distance::distribution}}
\label{app::serving::distance::distribtuion}
The conditional CCDF of $\|{\bf x}^*\|$ given ${\cal S}_i$ and ${\Phi}_{{\rm p}_k},\forall\ k\in{\cal K}_1$: is given by:
\begin{align*}
&\bar{F}_{s_i}(r|{\cal S}_i,\Phi_{{\rm p}_k}, \forall\ k\in{\cal K}_1) = \nbbP(\|{\bf x}^*\|>r|{\cal S}_i,\Phi_{{\rm p}_k}, \forall\ k\in{\cal K}_1) = \nbbP(\|\widetilde{\bf x}_i\|>r|{\cal S}_i,\Phi_{{\rm p}_k}, \forall\ k\in{\cal K}_1)\\&=\frac{\nbbP(\|\widetilde{\bf x}_i\|>r,{\cal S}_i|\Phi_{{\rm p}_k}, \forall\ k\in{\cal K}_1)}{\nbbP({\cal S}_i|\Phi_{{\rm p}_k}, \forall\ k\in{\cal K}_1)}  .
\end{align*}
Now
\begin{align*}
\nbbP(\|\widetilde{\bf x}_i\|>r,{\cal S}_i|\Phi_{{\rm p}_k}, \forall\ k\in{\cal K}_1)   =    \nbbP\left(\prod\limits_{j\in{\cal K}\setminus\{i\}}P_i \|\widetilde{{\bf x}}_i\|^{-\alpha}>P_j \|\widetilde{{\bf x}}_j\|^{-\alpha}, \|\widetilde{{\bf x}}_i\|>r| \Phi_{{\rm p}_k}, \forall\ k\in{\cal K}_1\right).
\end{align*}
This expression is similar to the  expression appearing in the computation of association probability with the additional event that $\|\widetilde{\bf x}_i\|>r$ which can be handled by changing the lower limit of the integral in \eqref{eq::association::probability}.  The final step is to differentiate the CCDF with respect to $r$. 
\subsection{Proof of Lemma~\ref{lemm::conditional::coverage}}
\label{app::conditional::coverage}
We first compute the probability of the $i^{th}$ coverage event given $\Phi_{{\rm p}_{k}},\forall\ k\in{\cal K}_1$ as follows. 
\begin{align*}
&\nbbP(\sinr({\bf x}^*)>\tau_i,{\cal S}_i|\Phi_{{\rm p}_k},\forall\ k\in{\cal K}_1) =\nbbP({\cal S}_i|\Phi_{{\rm p}_k},\forall\ k\in{\cal K}_1)\nbbP(\sinr({\bf x}^*)>\tau_i|{\cal S}_i,\Phi_{{\rm p}_k},\forall\ k\in{\cal K}_1),
\end{align*}where $\nbbP(\sinr({\bf x}^*)>\tau_i|{\cal S}_i,\Phi_{{\rm p}_k},\forall\ k\in{\cal K}_1)=$
\begin{align*}
& \nbbP\left(\frac{P_i h_{{\bf x}^*}\|{{\bf x}^*}\|^{-\alpha}}{{\tt N}_0+\sum\limits_{j\in{\cal K}}\sum\limits_{{\bf y}\in\Phi_{j}\setminus\{{\bf x}^*\}}P_j h_{{\bf y}}\|{{\bf y}}\|^{-\alpha}}>\tau_i\bigg|{\cal S}_i,\Phi_{{\rm p}_k},\forall\ k\in{\cal K}_1\right)\\
&= \nbbP\left( h_{{\bf x}^*}>\frac{\tau_i \|{\bf x}^*\|^{\alpha}}{P_i}\left({{\tt N}_0+\sum\limits_{j\in{\cal K}}\sum\limits_{{\bf y}\in\Phi_{j}\setminus\{{\bf x}^*\}}P_j h_{{\bf y}}\|{{\bf y}}\|^{-\alpha}}\right)\bigg|{\cal S}_i,\Phi_{{\rm p}_k},\forall\ k\in{\cal K}_1\right)\\
&\stackrel{(a)}{=}\nbbE\left[ \exp\left(-\frac{\tau_i \|{\bf x}^*\|^{\alpha}}{P_i}\left({{\tt N}_0+\sum\limits_{j\in{\cal K}}\sum\limits_{{\bf y}\in\Phi_{j}\setminus\{{\bf x}^*\}}P_j h_{{\bf y}}\|{{\bf y}}\|^{-\alpha}}\right)\right)\bigg|{\cal S}_i,\Phi_{{\rm p}_k},\forall\ k\in{\cal K}_1\right] \\
&= \nbbE\left[ \exp\left(-\frac{\tau_i {\tt N}_0 \|{\bf x}^*\|^{\alpha}}{P_i}\right)\prod\limits_{j\in{\cal K}}\prod\limits_{{\bf y}\in\Phi_{j}\setminus\{{\bf x}^*\}}\exp\left(-\frac{\tau_i P_j h_{{\bf y}}}{P_i}\frac{\|{{\bf x}^*}\|^{\alpha}}{\|{\bf y}\|^{\alpha}}\right)\bigg|{\cal S}_i,\Phi_{{\rm p}_k},\forall\ k\in{\cal K}_1\right] \\
&\stackrel{(b)}{=} \nbbE\left[ \exp\left(-\frac{\tau_i {\tt N}_0 \|{\bf x}^*\|^{\alpha}}{P_i}\right)\prod\limits_{j\in{\cal K}}\prod\limits_{{\bf y}\in\Phi_{j}\setminus\{{\bf x}^*\}}\nbbE_{h_{{\bf y}}}\left[\exp\left(-\frac{\tau_i P_j h_{{\bf y}}}{P_i}\frac{\|{{\bf x}^*}\|^{\alpha}}{\|{\bf y}\|^{\alpha}}\right)\right]\bigg|{\cal S}_i,\Phi_{{\rm p}_k},\forall\ k\in{\cal K}_1\right]\\
&\stackrel{(c)}{=}\nbbE\left[ \exp\left(-\frac{\tau_i {\tt N}_0 \|{\bf x}^*\|^{\alpha}}{P_i}\right)\prod\limits_{j\in{\cal K}}\prod\limits_{\substack{{\bf y}\in\Phi_{j}\cap \\b\left({\bf 0},\bar{P}_{j,i}\|{\bf x}^*\|\right)^c}}\frac{1}{1+\frac{\tau_i P_j}{P_i}\frac{\|{{\bf x}^*}\|^{\alpha}}{\|{\bf y}\|^{\alpha}}}\bigg|{\cal S}_i,\Phi_{{\rm p}_k},\forall\ k\in{\cal K}_1\right]\\
&=\int\limits_{0}^{\infty} \exp\left(-\frac{\tau_i {\tt N}_0 r^{\alpha}}{P_i}\right)\prod\limits_{j_1\in{\cal K}_1}\nbbE_{\Phi_{j_1}}\left[\prod\limits_{{\bf y}\in\Phi_{{j}_1}\cap b\left({\bf 0},\bar{P}_{j_1,i}r\right)^c}\frac{1}{1+\frac{\tau_i P_{j_1}}{P_i}\frac{r^{\alpha}}{\|{{\bf y}}\|^{\alpha}}}\bigg|\Phi_{{\rm p}_{j_1}}\right]\times\\
&\qquad\qquad\prod\limits_{j_2\in{\cal K}_2}\nbbE_{\Phi_{j_2}}\left[\prod\limits_{{\bf y}\in\Phi_{{j}_2}\cap b\left({\bf 0},\bar{P}_{j_2,i}r\right)^c}\frac{1}{1+\frac{\tau_i P_{j_2}}{P_i}\frac{r^{\alpha}}{\|{{\bf y}}\|^{\alpha}}}\right]f_{s_i}\left(r|{\cal S}_i,\Phi_{{\rm p}_k}, \forall\ k\in{\cal K}_1\right){\rm d}r\\
&= \int\limits_{0}^{\infty} \exp\left(-\frac{\tau_i {\tt N}_0 r^{\alpha}}{P_i}\right)\prod\limits_{j_1\in{\cal K}_1}\prod\limits_{{\bf z}\in\Phi_{{\rm p}_{j_1}}}
\exp\left(-\int\limits_{\bar{P}_{j_1,i} r}^{\infty} \bar{m}_{j_1} f_{{\rm d}_{j_1}}(y|{\bf z})\left(1-\frac{1}{1+\frac{\tau_i P_{j_1}}{P_i}\frac{{{r}}^{\alpha}}{y^{\alpha}}}\right){\rm d}y  \right)\times\\
&\qquad\qquad
\prod\limits_{j_2\in{\cal K}_2}
\exp
\left(
-2\pi\lambda_{j_2} \int\limits_{\bar{P}_{j_2,i} r}^{\infty} 
\left(
1-\frac{1}{1+\frac{\tau_i P_{j_2}}{P_i}\frac{{{ r}}^{\alpha}}{y^{\alpha}}}\right)
y{\rm d}y  
\right)
f_{s_i}\left(r|{\cal S}_i,\Phi_{{\rm p}_k}, \forall k\in{\cal K}_1\right){\rm d}r.\tag{\theequation}\label{myeq1}
\end{align*}
Here $(a)$ follows from the fact that $h_{{\bf x}^*}\sim\exp(1)$, $(b)$ follows from the assumption that all links undergo  i.i.d. fading, and $(c)$ is justified since in  $\Phi_{j}\setminus\{{\bf x}^*\}$, there exists no points inside the disc $b({\bf 0},\bar{P}_{j,i}\|{\bf x}^*\|)$, which is known as the {\em exclusion disc}.  
{In the last step, we use the PGFL of PPP from Lemma~\ref{lemm::pgfl::PPP}}. When $i\in{\cal K}_1$, substituting $f_{s_i}\left(r|{\cal S}_i,\Phi_{{\rm p}_k},\forall\ k\in{\cal K}_1\right)$ from \eqref{eq::serving::distance::distribution::PCP},  $\nbbP(\sinr({\bf x}^*)>\tau_i,{\cal S}_i|\Phi_{{\rm p}_k},\forall\ k\in{\cal K}_1)=$
\begin{align*}
&\bar{m}_{i}\int\limits_{0}^{\infty} \exp\left(-\frac{\tau_i {\tt N}_0 r^{\alpha}}{P_i}\right)\times\\&\prod\limits_{j_1\in{\cal K}_1}\prod\limits_{{\bf z}\in\Phi_{{\rm p}_{j_1}}}
\exp\left(-\bar{m}_{j_1}F_{{\rm d}_{j_1}}(\bar{P}_{j_1,i}r|z)-\bar{m}_{j_1}\int\limits_{\bar{P}_{j_1,i}r}^{\infty}  f_{{\rm d}_{j_1}}(y|{z})\left(1-\frac{1}{1+\frac{\tau_i P_{j_1}}{P_i}\frac{{{ r}}^{\alpha}}{y^{\alpha}}}\right){\rm d}y  \right)\times\\
&
\exp
\left(
-2\pi\sum\limits_{j_2\in{\cal K}_2}\lambda_{j_2} \int\limits_{\bar{P}_{j_2,i}r}^{\infty} 
\left(
1-\frac{1}{1+\frac{\tau_i P_{j_2}}{P_i}\frac{{{ r}}^{\alpha}}{y^{\alpha}}}\right)
y{\rm d}y  
-\pi\sum\limits_{j_2\in{\cal K}_2} \lambda_{j_2}\bar{P}_{j_2,i}^2 r^2
\right)\sum\limits_{{\bf z}\in\Phi_{{\rm p}_i}}f_{{\rm d}_i}(r|{z})
{\rm d}r \\
&=\bar{m}_{i}\int\limits_{0}^{\infty} \exp\left(-\frac{\tau_i {\tt N}_0 r^{\alpha}}{P_i}\right)\prod\limits_{j_1\in{\cal K}_1}\prod\limits_{{\bf z}\in\Phi_{{\rm p}_{j_1}}}
\exp\left(-\bar{m}_{j_1}\left(1-\int\limits_{\bar{P}_{j_1,i}r}^{\infty} \frac{ f_{{\rm d}_{j_1}}(y|{z})}{1+\frac{\tau_i P_{j_1}}{P_i}\frac{{{ r}}^{\alpha}}{y^{\alpha}}}{\rm d}y \right) \right)\times\\
&
\exp
\left(
-\pi r^2\sum\limits_{j_2\in{\cal K}_2}\lambda_{j_2}\bar{P}_{j_2,i}^2\tau_{i}^{2/\alpha} \int\limits_{\tau_i^{-2/\alpha}}^{\infty} 
\frac{1}{1+t^{\alpha/2}}{\rm d}t-\pi\sum\limits_{j_2\in{\cal K}_2} \lambda_{j_2}\bar{P}_{j_2,i}^2 r^2
\right)
\sum\limits_{{\bf z}\in\Phi_{{\rm p}_i}}f_{{\rm d}_i}(r|{z})
{\rm d}r.
\end{align*}
In the last step, we substitute $\frac{\tau_iP_{j_2}r^\alpha}{P_i y^\alpha} = t^{-\alpha/2}$. 
The final
expression is obtained by some algebraic simplification. 
 When $i\in{\cal K}_2$,  substituting  $f_{s_i}\left(r|\Phi_{{\rm p}_k}, k\in{\cal K}_1\right)$ in \eqref{myeq1} by \eqref{eq::serving::distance::distribution::PPP}, we get   $\nbbP(\sinr({\bf x}^*)>\tau_i,{\cal S}_i|\Phi_{{\rm p}_k},\forall\ k\in{\cal K}_1)=$ 
\begin{multline*}
 2\pi{\lambda}_{i}\int\limits_{0}^{\infty} \exp\left(-\frac{\tau_i {\tt N}_0 r^{\alpha}}{P_i}\right)\times\\\prod\limits_{j_1\in{\cal K}_1}\prod\limits_{{\bf z}\in\Phi_{{\rm p}_{j_1}}}
\exp\left(-\bar{m}_{j_1}F_{{\rm d}_{j_1}}(\bar{P}_{j_1,i}r|z)-\bar{m}_{j_1}
\int\limits_{\bar{P}_{j_1,i}r}^{\infty}  f_{{\rm d}_{j_1}}(y|{z})\left(1-\frac{1}{1+\frac{\tau_i P_{j_1}}{P_i}\frac{{{r}}^{\alpha}}{y^{\alpha}}}\right){\rm d}y  \right)\times\\
\exp
\left(
- \sum\limits_{j_2\in{\cal K}_2}2\pi \lambda_{j_2}\int\limits_{\bar{P}_{j_2,i}r}^{\infty} 
\left(
1-\frac{1}{1+\frac{\tau_i P_{j_2}}{P_i}\frac{{{r}}^{\alpha}}{y^{\alpha}}}
\right)
y{\rm d}y 
- \sum\limits_{j_2\in{\cal K}_2}\pi \lambda_{j_2}r^2 
 \bigg(\frac{P_{j_2}}{P_i}\bigg)^{2/\alpha}\right)r{\rm d}r.
\end{multline*}
The final
expression is obtained by some algebraic simplification. 
\subsection{Proof of Theorem~\ref{theorem::coverage::case::2}}
\label{app::theorem::coverage::case::2}
When $i\in {\cal K}_1\setminus\{q\}$, $\nbbP({\sinr}>\tau_i,{\cal S}_i)= $
\begin{multline*}
\bar{m}_{i}\int\limits_{0}^{\infty} \exp\left(-\frac{\tau_i {\tt N}_0 r^{\alpha}}{P_i}\right)\prod\limits_{j_1\in{\cal K}_1\setminus\{i,q\}}\nbbE_{\Phi_{{\rm p}_{j_1}}}\left[\prod\limits_{{\bf z}\in\Phi_{{\rm p}_{j_1}}}
{\cal C}_{j_1,i}(r,z)\right]\times\\
\nbbE_{\Phi_{{\rm p}_q}}\left[\prod\limits_{{\bf z}\in\Phi_{{\rm p}_{q}}}
{\cal C}_{q,i}(r,z)\right]
\exp
\left(
-\pi r^2\sum\limits_{j_2\in{\cal K}_2}\lambda_{j_2}\bar{P}_{j_2,i}^2\rho(\tau_i,\alpha)
\right)
\nbbE_{{\Phi_{{\rm p}_i}}}
\left[\sum\limits_{{\bf z}\in\Phi_{{\rm p}_i}}f_{{\rm d}_i}(r|{z})
\prod\limits_{{\bf z}\in\Phi_{{\rm p}_i}}{\cal C}_{i,i}(r,z)
\right]{\rm d}r.
\end{multline*}
As opposed to \caseS~1, the product over all points of $\Phi_{{\rm p}_q}$ has to be handled explicitly. The PGFL of $\Phi_{{\rm p}_q}$ for \caseS~2 evaluated at ${\cal C}_{q,i}(r,z)$ is given by
\begin{multline*}
\nbbE\left[\prod\limits_{{\bf z}\in \Phi_{{\rm p}_q}}{\cal C}_{q,i}(r,z) \right]  =\nbbE\left[{\cal C}_{q,i}(r,z_0) \right] \nbbE\left[\prod\limits_{{\bf z}\in \Phi(\lambda_{{\rm p}_q})}{\cal C}_{q,i}(r,z) \right] = 
\nbbE\left[{\cal C}_{q,i}(r,z_0)\right]\times\\\exp\bigg(-2\pi\lambda_{{\rm p}_q} \int\limits_{0}^{\infty}(1-{\cal C}_{q,i}(r,z))z{\rm d}{z}\bigg).
\end{multline*} 
This step is enabled by the fact that ${\bf z}_0$ and $\Phi(\lambda_{{\rm p}_q})$ are independent. 
Substituting $\mu({\bf z})= {\cal C}_{q,i}(z,r)$ and proceeding on similar lines of the proof of Theorem~\ref{theorem::coverage::case::1}, we obtain the final expression. Also note that since the function of ${\bf z}_0$ under consideration is only dependent on $\|{\bf z}_0\|\equiv z_0$, while taking expectation over ${\bf z}_0$, it is sufficient to consider the magnitude distribution of ${\bf z}_0$ which is denoted as $f_{{\rm d}_{\rm u}}(z_0|0)$. 
Now when $i=q$,
\begin{align*}
&\nbbP({\sinr}>\tau_i,{\cal S}_i)=\bar{m}_{q}\int\limits_{0}^{\infty} \exp\left(-\frac{\tau_q {\tt N}_0 r^{\alpha}}{P_q}\right)\prod\limits_{j_1\in{\cal K}_1\setminus\{q\}}\nbbE_{\Phi_{{\rm p}_{j_1}}}\left[\prod\limits_{{\bf z}\in\Phi_{{\rm p}_{j_1}}}
{\cal C}_{j_1,q}(r,z)\right]\times\\
&\exp
\left(
-\pi r^2\sum\limits_{j_2\in{\cal K}_2}\lambda_{j_2}\bar{P}_{j_2,q}^2\rho(\tau_q,\alpha)
\right)
\nbbE_{\Phi_{{\rm p}_{q}}}\left[\sum\limits_{{\bf z}\in\Phi_{{\rm p}_q}}f_{{\rm d}_q}(r|{z})
\prod\limits_{{\bf z}\in\Phi_{{\rm p}_q}}{\cal C}_{q,q}(r,z)
\right]{\rm d}r,
\end{align*}
where the last term in the expression under integral is the sum-product functional over $\Phi_{{\rm p}_{q}}$ which is computed as:
\begin{align*}
&\nbbE_{\Phi_{{\rm p}_{q}}}\left[\sum\limits_{{\bf z}\in\Phi_{{\rm p}_q}}f_{{\rm d}_q}(r|{z})
\prod\limits_{{\bf z}\in\Phi_{{\rm p}_q}}{\cal C}_{q,q}(r,z)
\right]  = 
\nbbE\left[f_{{\rm d}_{q}}(r|{z}_0){\cal C}_{q,q}(r,z_0)\right]\nbbE_{\Phi(\lambda_{{\rm p}_q})}\left[\prod\limits_{{\bf z}\in\Phi(\lambda_{{\rm p}_q})}{\cal C}_{q,q}(r,{z})\right]\\&+\nbbE_{\Phi(\lambda_{{\rm p}_q})}\left[\sum\limits_{{\bf z}\in{\Phi}_{{\rm p}_q}}f_{{\rm d}_q}(r|z)\prod\limits_{{\bf z}\in{\Phi}_{{\rm p}_q}}{\cal C}_{q,q}(r,{ z})\right]\nbbE[{\cal C}_{q,q}(r,{ z_0})]\\
&=\int\limits_0^{\infty}f_{{\rm d}_q}(r|z_0){\cal C}_{q,q}(r,z_0)f_{{\rm d}_{\rm u}}(z_0|0){\rm d}{z_0}\exp\left(-2\pi\lambda_{{\rm p}_q}\int\limits_{0}^{\infty}(1-{\cal C}_{q,q}(r,z)){z}{\rm d}z\right)\\&
+2\pi\lambda_{{\rm p}_q}\int\limits_{0}^{\infty}f_{{\rm d}_q}(r|z){\cal C}_{q,q}(r,z)z{\rm d}z\exp\left(-2\pi\lambda_{{\rm p}_q}\int\limits_{0}^{\infty}(1-{\cal C}_{q,q}(r,z)){z}{\rm d}z\right)\int\limits_{0}^{\infty}{\cal C}_{q,q}(r,z_0)f_{{\rm d}_{\rm u}}({z_0}|0){\rm d}z_0\\
&=\exp\left(-2\pi\lambda_{{\rm p}_q}\int\limits_{0}^{\infty}(1-{\cal C}_{q,q}(r,z)){z}{\rm d}z\right)\bigg(\int\limits_0^{\infty}f_{{\rm d}_q}(r|z_0){\cal C}_{q,q}(r,z_0)f_{{\rm d}_{\rm u}}(z_0|0){\rm d}{z_0}\\&
+2\pi\lambda_{{\rm p}_q}\int\limits_{0}^{\infty}f_{{\rm d}_q}(r|z){\cal C}_{q,q}(r,z)z{\rm d}z\int\limits_{0}^{\infty}{\cal C}_{q,q}(r,z_0)f_{{\rm d}_{\rm u}}({z_0}|0){\rm d}z_0\bigg).
\end{align*} 
The final expression can be obtained by proceeding on similar lines of the proof of Theorem~\ref{theorem::coverage::case::1}. 
Now, for $i\in{\cal K}_2$, $\nbbP({\sinr}>\tau_i,{\cal S}_i)= $
\begin{multline*}
2\pi\lambda_{i}\int\limits_{0}^{\infty} \exp\left(-\frac{\tau_i {\tt N}_0 r^{\alpha}}{P_i}\right)\prod\limits_{j_1\in{\cal K}_1\setminus\{q\}}\nbbE_{\Phi_{{\rm p}_{j_1}}}\left[\prod\limits_{{\bf z}\in\Phi_{{\rm p}_{j_1}}}
{\cal C}_{j_1,i}(r,z)\right]\times\\
\nbbE_{\Phi_{{\rm p}_{q}}}\left[\prod\limits_{{\bf z}\in\Phi_{{\rm p}_{q}}}
{\cal C}_{q,i}(r,z)\right]
\exp
\left(-\pi r^2\sum\limits_{j_2\in{\cal K}_2}\lambda_{j_2}\bar{P}_{j_2,i}^2\rho(\tau_i,\alpha)
\right)
r{\rm d}r.
\end{multline*}
Since expression is very similar to the expression of  $\nbbP({\sinr}>\tau_i,{\cal S}_i) $ obtained for $i\in{\cal K}_1\setminus\{q\}$,  we omit the next steps leading to the final expression. 
{\setstretch{1.20}
\bibliographystyle{IEEEtran}
\bibliography{MA_ref,MA_pub}

\begin{thebibliography}{10}
\providecommand{\url}[1]{#1}
\csname url@samestyle\endcsname
\providecommand{\newblock}{\relax}
\providecommand{\bibinfo}[2]{#2}
\providecommand{\BIBentrySTDinterwordspacing}{\spaceskip=0pt\relax}
\providecommand{\BIBentryALTinterwordstretchfactor}{4}
\providecommand{\BIBentryALTinterwordspacing}{\spaceskip=\fontdimen2\font plus
\BIBentryALTinterwordstretchfactor\fontdimen3\font minus
  \fontdimen4\font\relax}
\providecommand{\BIBforeignlanguage}[2]{{%
\expandafter\ifx\csname l@#1\endcsname\relax
\typeout{** WARNING: IEEEtran.bst: No hyphenation pattern has been}%
\typeout{** loaded for the language `#1'. Using the pattern for}%
\typeout{** the default language instead.}%
\else
\language=\csname l@#1\endcsname
\fi
#2}}
\providecommand{\BIBdecl}{\relax}
\BIBdecl

\bibitem{saha20173gpp}
C.~Saha, M.~Afshang, and H.~S. Dhillon, ``3{GPP}-inspired {H}et{N}et model
  using {P}oisson cluster process: Sum-product functionals and downlink
  coverage,'' \emph{IEEE Trans. on Commun.}, vol.~66, no.~5, pp. 2219--2234,
  May 2018.

\bibitem{HetHetNets2015}
M.~Mirahsan, R.~Schoenen, and H.~Yanikomeroglu, ``{HetHetNets}: Heterogeneous
  traffic distribution in heterogeneous wireless cellular networks,''
  \emph{IEEE Journal on Sel. Areas in Commun.}, vol.~33, no.~10, pp.
  2252--2265, Oct. 2015.

\bibitem{zhong2018effect}
Y.~Zhong, G.~Wang, R.~Li, T.~Han, X.~Ge, and T.~Q. Quek, ``Effect of spatial
  and temporal traffic statistics on the performance of wireless networks,''
  2018, available online: arxiv.org/abs/1804.06754.

\bibitem{3gppreportr12}
{3GPP TR 36.872 V12.1.0 }, ``3rd generation partnership project; technical
  specification group radio access network; small cell enhancements for
  {E-UTRA} and {E-UTRAN} - physical layer aspects (release 12),'' Tech. Rep.,
  Dec. 2013.

\bibitem{3gppreportr13}
{3GPP TR 36.932 V13.0.0 }, ``3rd generation partnership project; technical
  specification group radio access network; scenarios and requirements for
  small cell enhancements for {E-UTRA} and {E-UTRAN}; (release 13),'' Tech.
  Rep., Dec. 2015.

\bibitem{access2010further}
{3GPP TR 36.814}, ``Further advancements for {E-UTRA} physical layer aspects,''
  Tech. Rep., 2010.

\bibitem{AndrewsTractable}
J.~Andrews, F.~Baccelli, and R.~Ganti, ``A tractable approach to coverage and
  rate in cellular networks,'' \emph{IEEE Trans. on Commun.}, vol.~59, no.~11,
  pp. 3122--3134, 2011.

\bibitem{dhillon2012modeling}
H.~S. Dhillon, R.~K. Ganti, F.~Baccelli, and J.~G. Andrews, ``Modeling and
  analysis of {$K$}-tier downlink heterogeneous cellular networks,'' \emph{IEEE
  Journal on Sel. Areas in Commun.}, vol.~30, no.~3, pp. 550--560, Apr. 2012.

\bibitem{SahaAfshDh2016}
C.~Saha, M.~Afshang, and H.~S. Dhillon, ``Enriched {$K$}-tier {HetNet} model to
  enable the analysis of user-centric small cell deployments,'' \emph{IEEE
  Trans. on Wireless Commun.}, vol.~16, no.~3, pp. 1593--1608, Mar. 2017.

\bibitem{Mankar2016}
P.~D. Mankar, G.~Das, and S.~S. Pathak, ``Modeling and coverage analysis of
  {BS}-centric clustered users in a random wireless network,'' \emph{IEEE
  Wireless Commun. Letters}, vol.~5, no.~2, pp. 208--211, Apr. 2016.

\bibitem{AfshDhiClusterHetNet2016}
M.~Afshang and H.~S. Dhillon, ``Poisson cluster process based analysis of
  {HetNets} with correlated user and base station locations,'' \emph{IEEE
  Trans. on Wireless Commun.}, vol.~17, no.~4, pp. 2417--2431, Apr. 2018.

\bibitem{andrews2016primer}
J.~G. Andrews, A.~K. Gupta, and H.~S. Dhillon, ``A primer on cellular network
  analysis using stochastic geometry,'' 2016, available online:
  arxiv.org/abs/1604.03183.

\bibitem{elsawy2013stochastic}
H.~Elsawy, E.~Hossain, and M.~Haenggi, ``Stochastic geometry for modeling,
  analysis, and design of multi-tier and cognitive cellular wireless networks:
  A survey,'' \emph{IEEE Commun. Surveys Tuts.}, vol.~15, no.~3, pp. 996--1019,
  3th quarter 2013.

\bibitem{mukherjee2014analytical}
S.~Mukherjee, \emph{Analytical Modeling of Heterogeneous Cellular
  Networks}.\hskip 1em plus 0.5em minus 0.4em\relax Cambridge University Press,
  2014.

\bibitem{elsawy2016modeling}
H.~ElSawy, A.~Sultan-Salem, M.~S. Alouini, and M.~Z. Win, ``Modeling and
  analysis of cellular networks using stochastic geometry: A tutorial,''
  \emph{IEEE Commun. Surveys Tuts.}, vol.~19, no.~1, pp. 167--203, Firstquarter
  2017.

\bibitem{blaszczyszyn2018stochastic}
B.~B{\l}aszczyszyn, M.~Haenggi, P.~Keeler, and S.~Mukherjee, \emph{Stochastic
  geometry analysis of cellular networks}.\hskip 1em plus 0.5em minus
  0.4em\relax Cambridge University Press, 2018.

\bibitem{NonUniformDhillon}
H.~S. Dhillon, R.~K. Ganti, and J.~G. Andrews, ``Modeling non-uniform {UE}
  distributions in downlink cellular networks,'' \emph{IEEE Wireless Commun.
  Letters}, vol.~2, no.~3, pp. 339--342, Jun. 2013.

\bibitem{DownlinkChiranjib2016}
C.~Saha and H.~S. Dhillon, ``Downlink coverage probability of {$K$}-tier
  {HetNets} with general non-uniform user distributions,'' in \emph{Proc., IEEE
  Intl. Conf. on Commun. (ICC)}, May 2016.

\bibitem{TurgutUEclustering}
E.~Turgut and M.~C. Gursoy, ``Downlink analysis in unmanned aerial vehicle
  ({UAV}) assisted cellular networks with clustered users,'' \emph{IEEE
  Access}, vol.~6, pp. 36\,313--36\,324, May 2018.

\bibitem{andrews_lte_wifi}
Y.~Li, F.~Baccelli, J.~G. Andrews, T.~D. Novlan, and J.~C. Zhang, ``Modeling
  and analyzing the coexistence of {Wi-Fi} and {LTE} in unlicensed spectrum,''
  \emph{IEEE Trans. on Wireless Commun.}, vol.~15, no.~9, pp. 6310--6326, Sep.
  2016.

\bibitem{parida2017stochastic}
P.~Parida, H.~S. Dhillon, and P.~Nuggehalli, ``Stochastic geometry-based
  modeling and analysis of citizens broadband radio service system,''
  \emph{IEEE Access}, vol.~5, pp. 7326--7349, 2017.

\bibitem{Haenggi_gauss_poisson}
A.~Guo, Y.~Zhong, W.~Zhang, and M.~Haenggi, ``The {G}auss$-${P}oisson process
  for wireless networks and the benefits of cooperation,'' \emph{IEEE Trans. on
  Commun.}, vol.~64, no.~5, pp. 1916--1929, May 2016.

\bibitem{PairwiseTaylorDhillon2012}
D.~B. Taylor, H.~S. Dhillon, T.~D. Novlan, and J.~G. Andrews, ``Pairwise
  interaction processes for modeling cellular network topology,'' in
  \emph{Proc., IEEE Global Commun. Conf. (GLOBECOM)}, Dec. 2012, pp.
  4524--4529.

\bibitem{miyoshi2014cellular}
N.~Miyoshi and T.~Shirai, ``{A} cellular network model with {G}inibre
  configured base stations,'' \emph{Advances in Applied Probability}, vol.~46,
  no.~3, pp. 832--845, 2014.

\bibitem{nakata2014spatial}
I.~Nakata and N.~Miyoshi, ``Spatial stochastic models for analysis of
  heterogeneous cellular networks with repulsively deployed base stations,''
  \emph{Performance Evaluation}, vol.~78, pp. 7--17, 2014.

\bibitem{Li_Dhi_DPP}
Y.~Li, F.~Baccelli, H.~S. Dhillon, and J.~G. Andrews, ``Statistical modeling
  and probabilistic analysis of cellular networks with determinantal point
  processes,'' \emph{IEEE Trans. on Commun.}, vol.~63, no.~9, pp. 3405--3422,
  Sep. 2015.

\bibitem{DengHaenggiHeterogeneous2015}
N.~Deng, W.~Zhou, and M.~Haenggi, ``Heterogeneous cellular network models with
  dependence,'' \emph{IEEE Journal on Sel. Areas in Commun.}, vol.~33, no.~10,
  pp. 2167--2181, Oct. 2015.

\bibitem{yazdanshenasan2016poisson}
Z.~Yazdanshenasan, H.~S. Dhillon, M.~Afshang, and P.~H.~J. Chong, ``Poisson
  hole process: Theory and applications to wireless networks,'' \emph{IEEE
  Trans. on Wireless Commun.}, vol.~15, no.~11, pp. 7531--7546, Nov. 2016.

\bibitem{saha2017poisson}
C.~Saha, M.~Afshang, and H.~S. Dhillon, ``Poisson cluster process: {B}ridging
  the gap between {PPP} and {3GPP} {H}et{N}et models,'' in \emph{Proc.,
  Information Theory and Applications (ITA)}, 2017.

\bibitem{SuryaprakashMoller2015}
V.~Suryaprakash, J.M{\o}ller, and G.~Fettweis, ``On the modeling and analysis
  of heterogeneous radio access networks using a {P}oisson cluster process,''
  \emph{IEEE Trans. on Wireless Commun.}, vol.~14, no.~2, pp. 1035--1047, Feb.
  2015.

\bibitem{jo2012heterogeneous}
H.~S. Jo, Y.~J. Sang, P.~Xia, and J.~G. Andrews, ``Heterogeneous cellular
  networks with flexible cell association: A comprehensive downlink {SINR}
  analysis,'' \emph{IEEE Trans. on Wireless Commun.}, vol.~11, no.~10, pp.
  3484--3495, Oct. 2012.

\bibitem{AfshSahDhi2016Contact}
M.~Afshang, C.~Saha, and H.~S. Dhillon, ``Nearest-neighbor and contact distance
  distributions for {Thomas} cluster process,'' \emph{IEEE Wireless Commun.
  Letters}, vol.~6, no.~1, pp. 130--133, Feb. 2017.

\bibitem{AfshSahDhi2017ContactMatern}
------, ``Nearest-neighbor and contact distance distributions for {Mat{\'e}rn}
  cluster process,'' \emph{IEEE Commun. Letters}, vol.~21, no.~12, pp.
  2686--2689, Dec. 2017.

\bibitem{bao}
W.~Bao and B.~Liang, ``Handoff rate analysis in heterogeneous wireless networks
  with {P}oisson and {P}oisson cluster patterns,'' in \emph{Proc. 16th ACM
  International Symposium on Mobile Ad Hoc Networking and Computing}, ser.
  MobiHoc '15.\hskip 1em plus 0.5em minus 0.4em\relax New York, NY, USA: ACM,
  2015, pp. 77--86.

\bibitem{AzimiAbarghouyi2017StochasticGM}
S.~M. Azimi-Abarghouyi, B.~Makki, M.~Haenggi, M.~Nasiri-Kenari, and
  T.~Svensson, ``Stochastic geometry modeling and analysis of single- and
  multi-cluster wireless networks,'' \emph{IEEE Trans. on Commun.}, vol.~66,
  no.~10, pp. 4981--4996, Oct. 2018.

\bibitem{miyoshi2018downlink}
N.~Miyoshi, ``Downlink coverage probability in cellular networks with
  {P}oisson-{P}oisson cluster deployed base stations,'' \emph{IEEE Wireless
  Commun. Letters}, 2018, to appear.

\bibitem{chiu2013stochastic}
S.~N. Chiu, D.~Stoyan, W.~S. Kendall, and J.~Mecke, \emph{Stochastic Geometry
  and its Applications}, 3rd~ed.\hskip 1em plus 0.5em minus 0.4em\relax New
  York: John Wiley and Sons, 2013.

\bibitem{LoadAwareDhillon2013}
H.~S. Dhillon, R.~K. Ganti, and J.~G. Andrews, ``Load-aware modeling and
  analysis of heterogeneous cellular networks,'' \emph{IEEE Trans. on Wireless
  Commun.}, vol.~12, no.~4, pp. 1666--1677, Apr. 2013.

\bibitem{OffloadingSingh}
S.~Singh, H.~S. Dhillon, and J.~G. Andrews, ``Offloading in heterogeneous
  networks: Modeling, analysis, and design insights,'' \emph{IEEE Trans. on
  Wireless Commun.}, vol.~12, no.~5, pp. 2484--2497, May. 2013.

\bibitem{HaeInterferenceFucntionals}
U.~Schilcher, S.~Toumpis, M.~Haenggi, A.~Crismani, G.~Brandner, and
  C.~Bettstetter, ``Interference functionals in {P}oisson networks,''
  \emph{IEEE Trans. on Inf. Theory}, vol.~62, no.~1, pp. 370--383, Jan. 2016.

\bibitem{AfshDhi2015MehrnazD2D1}
M.~Afshang, H.~S. Dhillon, and P.~H.~J. Chong, ``Modeling and performance
  analysis of clustered device-to-device networks,'' \emph{IEEE Trans. on
  Wireless Commun.}, vol.~15, no.~7, pp. 4957--4972, Jul. 2016.

\bibitem{afshang2018equi}
M.~Afshang, C.~Saha, and H.~S. Dhillon, ``Equi-coverage contours in cellular
  networks,'' \emph{IEEE Wireless Commun. Letters}, vol.~7, no.~5, pp.
  700--703, Oct. 2018.

\bibitem{saha2018IAB}
C.~Saha, M.~Afshang, and H.~S. Dhillon, ``Bandwidth partitioning and downlink
  analysis in millimeter wave integrated access and backhaul for 5{G},''
  \emph{IEEE Trans. on Wireless Commun.}, 2018, to appear.

\end{thebibliography}
}
\end{document}